\input jytex.tex   % available from hep-th
%\draft
\typesize=10pt \magnification=1200
\baselineskip17truept %\baselineskip25truept
\hsize=6truein\vsize=8.5truein %\leftmargin=1.25in
%\oddleftmargin=.5in
%\evenleftmargin=1.5in
\sectionnumstyle{blank}
\chapternumstyle{blank}
\chapternum=1
\sectionnum=1
\pagenum=0
\input epsf.sty
%\referencestyle{preordered}
% title style follows

\def\begintitle{\pagenumstyle{blank}\parindent=0pt\begin{narrow}[0.4in]}
\def\endtitle{\end{narrow}\newpage\pagenumstyle{arabic}}

% exercise style follows

\def\beginexercise{\vskip 20truept\parindent=0pt\begin{narrow}[10
truept]}
\def\endexercise{\vskip 10truept\end{narrow}}

% **************    my jyTeX abbreviations   *****************

\def\eql#1{\eqno\eqnlabel{#1}}
\def\ref{\reference}
\def\peq{\puteqn}
\def\pref{\putref}

\def\mgn{\marginnote}
\def\bex{\begin{exercise}}
\def\eex{\end{exercise}}

% *********************** My definitions ************************

\font\open=msbm10 %scaled\magstep1 % For VAX. Borde p195.
 %scaled\magstep1 % For VAX. Borde p195.
%\font\open=msym10 %scaled\magstep1 % For Arbortxt on PC
%\font\opens=msym8 %scaled\magstep1 % For Arbortxt on PC
%\font\goth=eufm10  % For Arbortxt on PC, and VAX. Borde p199
%\font\ssb=cmss10
%\font\smsb=cmss8
\def\mbox#1{{\leavevmode\hbox{#1}}}

\def\hspace#1{{\phantom{\mbox#1}}}
\def\oR{\mbox{\open\char82}}

\def\oZ{\mbox{\open\char90}}

\def\rd{{\rm d}}

\def\rS{{\rm S}}

\def\al{\alpha}
 %in jyTeX
 %in jyTeX
 %in jyTeX
 %in jyTeX
 %in jyTeX
 %in jyTeX
% in jyTeX
% in jyTeX
% in jyTeX
\def\be{\beta}
\def\ga{\gamma}
\def\de{\delta}
\def\Ga{\Gamma}

\def\la{\lambda}
\def\La{\Lambda}
\def\om{\omega}
\def\Om{\Omega}

\def\si{\sigma}

\def\th{\theta}

\def\ze{\zeta}

\def\Real{{\rm Re\,}}

\def\Res{{\rm Res\,}}
\def\zf{$\zeta$--function}
\def\zfs{$\zeta$--functions}

     % Newline

\def\frac#1/#2{\leavevmode\kern.1em
\raise.5ex\hbox{\the\scriptfont0 #1}\kern-.1em/\kern-.15em
\lower.25ex\hbox{\the\scriptfont0 #2}}
\def\sfrac#1/#2{\leavevmode\kern.1em
\raise.5ex\hbox{\the\scriptscriptfont0 #1}\kern-.1em/\kern-.15em
\lower.25ex\hbox{\the\scriptscriptfont0 #2}}
\def\half{{1\over 2}}
\def\gtorder{\mathrel{\raise.3ex\hbox{$>$}\mkern-14mu
             \lower0.6ex\hbox{$\sim$}}}
\def\ltorder{\mathrel{\raise.3ex\hbox{$<$}\mkern-14mu
             \lower0.6ex\hbox{$\sim$}}}

\def\semidirprod{\rlap{\ss C}\raise1pt\hbox{$\mkern.75mu\times$}}
\def\for{\lower6pt\hbox{$\Big|$}}
\def\fish{\kern-.25em{\phantom{abcde}\over \phantom{abcde}}\kern-.25em}

 %triple
%dot
 %double
%dot
 %double dot
%for small #1

\def\boxit#1{\vbox{\hrule\hbox{\vrule\kern3pt
        \vbox{\kern3pt#1\kern3pt}\kern3pt\vrule}\hrule}}
\def\dalemb#1#2{{\vbox{\hrule height .#2pt
        \hbox{\vrule width.#2pt height#1pt \kern#1pt
                \vrule width.#2pt}
        \hrule height.#2pt}}}

        %double stroke
\def\frac#1#2{{{#1}\over{#2}}}
 %lower covariant deriv.
    %lower ordinary  deriv.

\def\pafra#1#2{{\pa#1\over\pa#2}}

      %Connection
    %Connection'
\def\comb#1#2{{\left(#1\atop#2\right)}}

\def\etc{{\it etc. }}

\def\eg{{\it e.g. }}
\def\ie{{\it i.e. }}
\def\cf{{\it cf }}
\def\pa{\partial}
\def\bra#1{\langle#1\mid}

 %gives average <#1>
 %gives thermal average <<#1>>
\def\br#1#2{\langle{#1}\mid{#2}\rangle}   %gives bracket <#1|#2>
 %gives big bracket <#1|#2>
  %gives
%matrix element <#1|#2|#3>

\def\Tr{{\rm Tr\,}}

\def\wr{{\widehat{\bf r}}}
\def\br{{\bf r}}
\def\bv#1{{\bf #1}}

\def\3j#1#2#3#4#5#6{\left\lgroup\matrix{#1&#2&#3\cr#4&#5&#6\cr}
\right\rgroup}

\def\mod{{{\rm S}^2/\Gamma}}

\def\caI{{\cal I}}

\def\caP{{\cal P}}
\def\caY{{\cal Y}}

\def\caF{{\cal F}}
\def\caD{{\cal D}}
\def\man{{\cal M}}

\def\m?{\mgn{?}}

% KK's defs

\def\pa{\partial}

\def\beq{\begin{eqnarray}}
\def\eeq{\end{eqnarray}}

%  *******************  Journal refs **********************

\def\aop#1#2#3{{\it Ann. Phys.} {\bf {#1}} (19{#2}) #3}

\def\cmp#1#2#3{{\it Comm. Math. Phys.} {\bf {#1}} (19{#2}) #3}
\def\cqg#1#2#3{{\it Class. Quant. Grav.} {\bf {#1}} (19{#2}) #3}
\def\hpa#1#2#3{{\it Helv.Phys.Acta} {\bf {#1}} (19{#2}) #3}

\def\ijtp#1#2#3{{\it Int. J. Theor. Phys.} {\bf {#1}} (19{#2}) #3}
\def\jgp#1#2#3{{\it J. Geom. and Phys.} {\bf {#1}} (19{#2}) #3}
\def\jmp#1#2#3{{\it J. Math. Phys.} {\bf {#1}} (19{#2}) #3}
\def\jpa#1#2#3{{\it J. Phys.} {\bf A{#1}} (19{#2}) #3}

\def\np#1#2#3{{\it Nucl. Phys.} {\bf B{#1}} (19{#2}) #3}
\def\pl#1#2#3{{\it Phys. Lett.} {\bf {#1}} (19{#2}) #3}

\def\physica#1#2#3{{\it Physica} {\bf {#1}} (19{#2}) #3}

\def\prD#1#2#3{{\it Phys. Rev.} {\bf D{#1}} (19{#2}) #3}

\def\zfp#1#2#3{{\it Z. f. Phys.} {\bf {#1}} (19{#2}) #3}

\def\cras#1#2#3{{\it Comptes Rend. Acad. Sci. (Paris)} {\bf{#1}} (#2) #3}

\def\mpcps#1#2#3{{\it Math. Proc. Camb. Phil. Soc.} {\bf{#1}} (19{#2}) #3}

\def\am#1#2#3{{\it Acta Mathematica} {\bf {#1}} (19{#2}) #3}
\def\aim#1#2#3{{\it Adv. in Math.} {\bf {#1}} (19{#2}) #3}
\def\ajm#1#2#3{{\it Am. J. Math.} {\bf {#1}} ({#2}) #3}

\def\aom#1#2#3{{\it Ann. of Math.} {\bf {#1}} (19{#2}) #3}
\def\cjm#1#2#3{{\it Can. J. Math.} {\bf {#1}} (19{#2}) #3}
\def\cpde#1#2#3{{\it Comm. Partial Diff. Equns.} {\bf {#1}} (19{#2}) #3}

\def\invm#1#2#3{{\it Invent. Math.} {\bf {#1}} (19{#2}) #3}
\def\ijpam#1#2#3{{\it Ind. J. Pure and Appl. Math.} {\bf {#1}} (19{#2}) #3}
\def\jdg#1#2#3{{\it J. Diff. Geom.} {\bf {#1}} (19{#2}) #3}

\def\jmpa#1#2#3{{\it J. Math. Pures. Appl.} {\bf {#1}} ({#2}) #3}

\def\ojm#1#2#3{{\it Osaka J.Math.} {\bf {#1}} ({#2}) #3}

\def\pja#1#2#3{{\it Proc. Jap. Acad.} {\bf {A#1}} (19{#2}) #3}

\def\tams#1#2#3{{\it Trans. Am. Math. Soc.} {\bf {#1}} (19{#2}) #3}

% *******************   Main text *********************
%\begin{ignore}
\begin{title}
\vglue 1truein
%\righttext {ICTP/2?}
%\righttext{hep-th/98}
\vskip15truept
%\leftline{\today}
%\vskip 30truept
\centertext {\Bigfonts \bf Magnetic fields and factored two-spheres}
\vskip 20truept
\centertext{J.S.Dowker\footnote{{Permanent address: Department of
Theoretical Physics,
The University of Manchester, Manchester, England}}}
\centertext{{\it Theory Group, Department of Physics, Imperial College,}}
\centertext{{\it Blackett Laboratory, Prince Consort Rd, London.}}
\vskip 20truept
\centertext {Abstract}
\vskip10truept
\begin{narrow}
A magnetic monopole is placed at the centre of a ball whose surface
$\rS^2$ is tiled by the symmetry group, $\Ga$, of a regular solid. The
quantum mechanics on the two-dimensional quotient $\rS^2/\Ga$ is developed
and the monopole charge is found to be quantised in an expected manner.
The heat-kernel and \zfs\ are evaluated and the Casimir energy is computed.
Numerical approaches to the calculation of the derivative of the Barnes
\zf\ are presented.
\end{narrow}
\vskip 5truept
\righttext {June 1999}
\vskip 60truept
%\righttext{Typeset in \jyTeX}
\vfil
\end{title}
\pagenum=0
%\end{ignore}
\section{\bf 1. Introduction}

In previous work, [\pref{ChandD}], we have discussed quantum field theory on
orbifold factors of spheres ${\rm S}^n/\Ga$ where $\Ga$ is a discrete
subgroup of the orthogonal group, ${\rm O}(n+1)$.
In this paper we wish to present an extension
in the two-dimensional case ($n=2$) to the situation where there is a uniform
radial magnetic field passing through the surface of the sphere. This can be
thought of as due to a magnetic monopole at the centre of a ball in an
embedding ${\rm R}^3$. The motivation is partly to investigate the interplay
between a magnetic field and a non-trivial topology/geometry induced by
identification. Specifically we would be interested in what happens to
the Dirac quantization in topologically interesting or singular manifolds
(orbifolds). There is also continuing statistical mechanical interest in
magnetic fields and two-dimensional domains.

The quotient group is the complete symmetry group of a regular solid and
can be generated by reflections in the three (concurrent) planes of symmetry
of the solid. The elements fall into two sets depending on whether they
contain an even or an odd number of relections. The even subset forms the
rotational subgroup denoted now by $\Ga\in {\rm{SO}}(3)$. The odd subset
is denoted by $\Ga_1'$. If $\ga'$ is any fixed element of $\Ga'$ then
as $\ga$ runs over $\Ga$, $\ga\ga'$ exhausts $\Ga'_1$.
In particular we can choose $\ga'=\si$ where  $\si$ is a reflection
in a symmetry plane and so the complete group is
$$\Ga'=\Ga\cup\Ga'_1=\Ga\cup\Ga\si.
\eql{gdecomp}$$

The standard classification of finite subgroups (reflection groups)
is given by Meyer, [\pref{Meyer}], and we use his notation.
The general construction of the quotients, $\rS^2/\Ga'$, which in this case
are certain geodesic triangles on $\rS^2$ is sketched later.
The vertices of these triangles are singular points. Joining them to the
origin of the ambient $\oR^3$ produces singular strings and our analysis
can be extended to this three-dimensional setting [\pref{ChandD2}]. The
two-dimensional models that we study can be thought of as toy models
for more general `textures' in higher dimensions as laid down by
Kibble, [see \eg [\pref{CS}]).

Our main interest is in setting up the general framework and then
applying it to some specific calculations such as the evaluation
of vacuum energies. This will involve a certain amount of
technical manipulation, especially with the properties of the
Barnes \zf. This function appears quite commonly, particularly
in spherically symmetric situations and in the presence of
magnetic fields or harmonic oscillators and we expect that our
methods will have an applicability beyond the immediate one here.

A very brief summary of our findings is given in section 12.

\section{\bf 2. Modes and group actions on the full sphere.}

As modes on the full sphere, we can take the angular part of the
Schr\"odinger equation solutions
derived long ago by Tamm [\pref{Tamm}] and Fierz [\pref{Fierz}]. It would be
 more rigorous to use a fibre
bundle formulation (Greub and Petry [\pref{GandP}] , Wu and Yang
[\pref{WandY}]) or
geometric quantisation [\pref{Snia,Horvathy}] but this is unnecessary for our
purposes. For definiteness, we employ the modes denoted by $(Y_{qm})_a$
in Wu and Yang, corresponding to the string running down the negative
$z$--axis. The modes are, up to normalisation, the ${\rm SU}(2)$
representation matrices $\caD^{(l)*}_{m,-q}(\phi,\th,-\phi)$
with $l=|q|,|q+1|\ldots,\,-l\le m \le l$ [\pref{BandL}]). $2q$ is the
monopole number with $q\equiv eg/\hbar$, and $2q\in \oZ$. If the string runs
 up
the positive $z$-axis, the modes are $\caD^{(l)*}_{m,-q}(\phi,\th,\phi)$.

The corresponding eigenvalues of $H_{\rS^2}$, the angular part of
$-(\nabla-ie{\bf A})^2$, are
$$
\la_l=l(l+1)-q^2=\bigg(l+{1\over2}\bigg)^2-{1\over4}-q^2
\eql{eigenv}
$$
with degeneracy $2l+1$.

It is helpful to give the explicit form of the angular eigenfunctions in
spherical polar coordinates,
$$
Y^{(l)}_{qm}(\theta,\phi)=N_{qlm}\sin^{|q+m|}(\theta/2)
\cos^{|q-m|}(\theta/2)P^{|q+m|,|q-m|}_{l-(|q+m|+|q-m|)/2}(\cos\theta)
e^{i(q+m)\phi}
\eql{mon:harm1}$$
where $N_{qlm}$ is a normalisation constant and the $P^{\alpha,\beta}_n(x)$
are
Jacobi polynomials. In the Wu-Yang formalism these are the
solutions (sections) in the upper hemisphere. In the lower
hemisphere the potential is taken to have a string along the
positive $z$-axis. The two sets of solutions are related on the
equator by the factor $\exp(i2q\phi)$. Making this single valued
gives the quantisation condition on $q$, in this approach.

The basic means of finding the modes is to write
$H_{\rS^2}\psi=\lambda\psi$ explicitly as a differential equation in
spherical
polar coordinates,
$$
{1\over\sin\theta}\pafra{}{\theta}\sin\theta\pafra{\psi}{\theta}+
{1\over\sin^2\theta}\pafra{^2\psi}{\phi^2}+i
{2q\over 1+\cos\theta}\pafra{\psi}{\phi}-
q^2{1-\cos\theta \over 1+\cos\theta}\psi=\lambda\psi\,,
\eql{mon:diff}$$
and solve it assuming the
separation\break $\psi(\theta,\phi)=\exp(iu\phi)P(\cos\theta)$. Using this
method we can suppose
initially that $q$ is arbitrary which leads to the same solutions as in
(\peq{mon:harm1}) but which are characterised by two integers $u\in\oZ^+$ and
$v\in\oZ$, [\pref{Tamm}]. The relationship to the $SU(2)$ labels $l$ and $m$
is given by
$$
l=u+\half\big(|q+m|+|q-m|\big),\quad v=m-q
\eql{mon:harm2}$$
and thus it would appear that we could have any value of $q$ as long as $u$
 and
$v$ are integers (with suitable $l,\,m\in\oR$). However, in this case the
solutions would not be single-valued and all the solutions would vanish on
the
string axis. Setting $2q\in\oZ$ gives the same values of $l$ and $m$ as
before.

As usual, when one tries to adapt wave-functions to some symmetry (here
$\Ga'$) there is the problem that the magnetic potential, ${\bf A}$, might
not possess the same symmetry so that compensating gauge transformations
are necessary. Peierls, [\pref{Peierls}], calls this process `umeichen'.
It is a well
known situation, with extensive discussion,[\pref{JandJ}], which we have
 encountered in
an earlier calculation on the tetrahedron, [\pref{tetra}], an orbifold
factor of the plane.
A similar procedure has also been applied to factors of the Poincar\'e
half-plane.

In the present case, under the action of $\Ga'$, the string is
rotated and reflected and has to be brought back to its original position
if we are to implement the identification in ${\rm
S}^2/\Ga'$ consistently. The equation that contains the relevant
information is the behaviour of the modes under arbitrary
rotations-reflections. For the moment we deal with the easier,
pure rotational case, $g\in {\rm SO}(3)$. Then the behaviour is an
elementary consequence of the ${\rm SU}(2)$ group combination law
and one has explicitly (Wu and Yang [\pref{WandY}], Frenkel and
Hrask\'o [\pref{FandH}])
$$
Y^{(l)}_{qm'}(g^{-1}\wr)=e^{iq\La_g(\hat\br)}\,
Y^{(l)}_{qm}(\wr)\,\caD^{(l)}_{mm'}(g),\quad g\in {\rm SO}(3)\,.
\eql{modrot}$$
The exponential factor is the compensating gauge
rotation with $$ \La_g(\wr)=\al(g)+\ga(g)-\Om_g \eql{phase}$$
where $\al$, $\be$ and $\ga$ are the Euler angles of the rotation
$g$ and $\Om_g$ is the solid angle subtended by the geodesic
triangle on the (unit) sphere cut out by $-{\bf r}$, the string
axis, ${\bf n}$, (here the negative $z$-axis) and the rotated
string axis, $g{\bf n}$, $$ \Om_g(\wr)=\Om(g{\bf n},{\bf n};{\bf
r}),\quad {\bf r}=(\wr,r). $$ (Our conventions regarding rotations
and actions are generally those of Brink and Satchler [\pref{BandS}].)The gradient
of $\Om$ gives the gauge transformation between the potentials of the
two strings.

An alternative expression for $\La$ is (Wu and Yang [\pref{WandY}])
$$
\La_g(\wr)=\phi_g-\phi-A_g
\eql{phase2}$$
where $A_g$ is the angle at $-\wr$ in the above mentioned triangle on
$\rS^2$ and $g^{-1} \wr=(\th_g,\phi_g)$

In order that (\peq{modrot}) be consistently iterated, it is necessary
that the group combination (cocycle) condition,
$$
\La_{gh}(\wr)=\La_g(\wr)+\La_h(g^{-1}\wr),
\eql{coc}$$
should hold, and this can be checked directly from (\peq{phase}). The
geometrical details are to be found in the useful article by Frenkel and
Hrask\'o [\pref{FandH}].

The magnetic rotation operator, $T_g$, is defined, on scalars, by the
action
$$
(T_g\psi)({\bf r})=e^{-iq\La_g(\,\wr\,)}\,\psi(g^{-1}{\bf r}),
\eql{magop}$$
or, equivalently,
$$
\bra{{\bf r}}T_g=e^{-iq\La_g(\,\wr\,)}\,\bra{g^{-1}{\bf r}}\,.
\eql{magop2}$$
In the present spherical case, because of (\peq{coc}), the magnetic
rotations provide a true, as opposed to a ray,
representation of the {\it double} of ${\rm SO}(3)$
(denoted ${\rm SO}^\bullet(3)$) in the sense that
$$
T_{gh}=T_g\,T_h
$$
and
$$
T_E=(-1)^{2q}{\bf 1}\,,
\eql{2pirot}$$
where $E$ is a $2\pi$ rotation.
This is not unexpected considering that for $q$ a half odd-integer all
the monopole harmonics have half odd-integer angular momentum.
${\rm SO}^\bullet(3)$ is isomorphic to ${\rm SU}(2)$.

Magnetic rotations on the plane [\pref{Tam}] also provide true
representations but not so
{\it translations}, unless the flux through a fundamental domain is
quantised [\pref{OandT}].

We record the action of $T_g$ on the modes which is easily obtained from
(\peq{magop}) and (\peq{modrot}),
$$
T_g Y^{(l)}_{qm'}(\wr)=Y^{(l)}_{qm}(\wr)\,\caD^{(l)}_{mm'}(g).
\eql{modact}$$
\section{\bf 3. Parity and reflections.}
The singularity (string) preserving magnetic parity operator on monopole
wavefunctions is defined by the action, [\pref{FandH}],
$$
(\Pi\psi)({\bf r})=e^{-iq\Om(-{\bf n},{\bf n};{\bf r})}\,\psi(-{\bf r})
\eql{parity},$$
under the inversion $\iota:{\bf r}\to -{\bf r}$

On the modes we have, as an easy calculation shows,
$$
\Pi Y^{(l)}_{qm}=(-1)^l\,Y^{(l)}_{-qm}.
\eql{modpar}$$

The parity operator can be used to extend the magnetic rotation operator
$T$ to include reflections. The reflection, $\si$, in the plane with
normal ${\bf t}$ can be written as a rotation through $\pi$ about the
axis ${\bf t}$ combined with the parity inversion $\iota$. This can be
written as $\si=g_\pi\,\iota=\iota\,g_\pi$. The extension of $T$ to
reflections is thus defined by
$$
T_{\si}\psi=T_{\iota g_\pi}\psi=T_{\iota}\,T_{g_\pi}\psi=\Pi\,T_{g_\pi}
\psi.
$$
From (\peq{magop}) and (\peq{parity})
$$
(T_{\si}\psi)({\bf r})=e^{-iq\Om_{\si}(\hat\br)+iq\pi}\,(R_{g_\pi}\psi)
(-{\bf r}),
\eql{reflecop}$$
and on the modes,
$$
(T_{\si}Y^{(l)}_{qm'})(\wr)=(-1)^l\,Y^{(l)}_{-qm}(\wr)\,\caD^{(l)}_{mm'}
(g_\pi).
\eql{reflact}$$

\section{\bf 4. Construction of the domains $\rS^2/\Ga'$}
It is convenient now to formalize briefly what we mean by
the space ${\rm S}^2/\Ga'$ where $\Ga'$ is a finite subgroup of
$O(3)$. A region of the sphere $\caF\in\rS^2$ is called a
fundamental domain for $\Ga$ if it satisfies the following
criteria,

(i) $\caF$ is open in $\rS^2$ ,

(ii) $\caF \bigcap \ga\caF=\emptyset,\quad \forall
\ga\in\Ga'-\{id\}$,

(iii) $\rS^2=\bigcup_{\ga\in\Ga'}\overline{\ga\caF}$.

\noindent We also assume that $\caF$ is connected.

Physically, $\caF\bigcup\partial\caF$ represents the space on
which our theory is defined, and all the copies of $\caF$ on
$\rS^2$ must be physically equivalent.

The space $\rS^2/\Ga'$ is the the sphere $\rS^2$ with the points $\bf r$
and $\ga\bf r$ identified for all $\ga\in\Ga'$. If $\Ga'$ acts freely on
$\rS^2$, \ie there are no fixed points, then $\rS^2/\Ga'$ is closed and
can be taken to be $\caF$ with identified boundary. If there are fixed
points, these will be contained in $\partial\caF$ and can be considered
as boundary singular points of $\rS^2/\Ga'$.

When $\Ga$ is purely rotational the fixed points form a discrete set of
which there are two, or three, in $\pa\caF$. For the extended, reflection
groups, as already indicated, the boundary $\pa\caF$ is constructed
from the intersections of two or three reflection planes with the sphere.
Thus there are transformations that map $\caF$ into adjacent domains
and which leave part of the boundary of $\caF$ fixed. The conclusion here
is that, unlike the rotational case, the boundary $\pa\caF$ is a real
boundary and the physical manifold is termed a M\"obius triangle.

\section {\bf 5. Projection to ${\rm S}^2/\Ga$}

Still restricting to purely rotational $\Ga$, we will now formally
project everything down to ${\rm S}^2/\Ga$ in a more or less
standard fashion.

In order that a function $\widetilde\psi(\br)$ on $\rS^2$ project
down consistently to ${\rm S}^2/\Ga$, it is necessary that it satisfy the
magnetic periodicity condition
$$T_{g^\bullet}\widetilde\psi(\br)=a(g^\bullet)
\widetilde\psi(\br),\quad g\in\Ga^\bullet,
\eql{period} $$
where $a(g^\bullet)$ forms a one-dimensional
representation of $\Ga^\bullet$, the double of $\Ga$, and labels the
projection. For ease we have put $\br$ in place of $\wr$. We show
later that there is a minimal choice for $a(g^\bullet)$.

If $q$ is integral, there is no need to introduce the group doubles.
However, if one does, there is a simple duplication, which is
easily dealt with in practice as we can show with the following,
somewhat superfluous, constructions.

Since the action of $\Ga^\bullet$ covers the sphere $\rS^2$ twice, we
introduce the trivial double covering, ${\rS^2}^\bullet$, of two
identical copies of $\rS^2$, with
${\rS^2}^\bullet/\Ga^\bullet={\rS^2/\Ga}$.

The modes themselves do not change sign on a $2\pi$ rotation,
even when $l$ is half odd-integral, and neither does the
wavefunction. So $\psi^\bullet(\br)=\psi^\bullet(E\br)$, where $E$
is a $2\pi$ rotation, and therefore,
$\widetilde\psi(\br)=\psi^\bullet(\br)$.

We then have, somewhat non-rigorously, for the function,
$\psi(\br)$, on ${\rS^2}^\bullet/\Ga^\bullet$ the numerical equality
$$
\psi(\br)=\psi^\bullet(\br)
\eql{nequal}$$
where we do not distinguish
between the projected and original coordinates, both being denoted
by $\br$. The $\br$ on the right belongs to the fundamental domain
of $\Ga^\bullet$ on ${\rS^2}^\bullet$, which is, of course, isomorphic
to that of $\Ga$ on $\rS^2$. Then
$$T_{g^\bullet}\psi(\br)=a(g^\bullet)\psi(\br). \eql{period2}
$$

The heat-kernel on ${\rm S}^2$, written in mode form,
$$
K_{\rS^2}\big(\br,\br';\tau\big)=\sum_{l=|q|}^\infty
e^{-\la_l\tau} \Tr \left[{\bf Y}_q^{(l)}(\br){\bf Y}_q^{(l)
\dag}(\br')\right],
\eql{kernel}$$
propagates arbitrary wavefunctions  on $\rS^2$ and
satisfies the important switching relation
$$
K_\rS\big(\br,g^{-1}\br';\tau\big)=e^{-iq\La_g(\br')-iq\La_{g^{-1}}(\br)}\,
K_\rS\big(g\br,\br';\tau\big) $$
which realises in coordinate representation
the operator rotational symmetry
$$ T_g K_{\rS^2}=K_{\rS^2}T_g
$$
(see (\peq{magop2})).

The heat-kernel that propagates wavefunctions  on
${\rS^2}^\bullet/\Ga^\bullet$ obeying  (\peq{period2}) is, by
general theory,
$$
K_{{\rS^2}^\bullet/\Ga^\bullet}(\br,\br')=\sum_{g\in\Ga^\bullet}
a(g^\bullet)\,T_{{g^\bullet}^{-1}}\,K_{{\rS^2}^\bullet}(\br,\br')
\eql{pkernel}$$
which is referred to as the pre-image form of this propagator in
terms of that on ${\rS^2}^\bullet$.

The double group $\Ga^\bullet$ can be decomposed
$$
\Ga^\bullet=\Ga\cup E\Ga
$$
and the sum over $\Ga^\bullet$ reduced to a sum over the subset $\Ga$
$$
K_{{\rS^2}^\bullet/\Ga^\bullet}(\br,\br')=\sum_{g\in\Ga}
\big(a(g)\,T_{{g}^{-1}}+a(Eg)\,T_{(Eg)^{-1}}\big)
\,K_{{\rS^2}^\bullet}(\br,\br').
\eql{pkernel2}$$
We know, (\peq{2pirot}), that $T_E=(-1)^{2q}{\bf1}$ and so from
(\peq{period2}), $a(E)=(-1)^{2q}$. Hence $a(E)\,T_E={\bf1}$ and so
$$
K_{{\rS^2}^\bullet/\Ga^\bullet}(\br,\br')=2\sum_{g\in\Ga}
a(g)\,T_{{g}^{-1}}
\,K_{{\rS^2}^\bullet}(\br,\br'),
\eql{pkernel3}$$
which means we can finally write what was almost obvious from the start, the
preimage sum,
$$
K_{\rS^2/\Ga}(\br,\br')=\sum_{g\in\Ga}
a(g)\,T_{g^{-1}}
\,K_{\rS^2}(\br,\br').
\eql{pkernel4}$$
The factor of 2 is a normalization (volume) factor between
${\rS^2}^\bullet$ and $\rS^2$, $K_{{\rS^2}^\bullet}=2K_{\rS^2}$.

This result shows that we can effectively ignore the complication of the
double group and just proceed with $\Ga$ as usual.

Given an arbitrary field $\widetilde\phi(\br)$ on $\rS^2$, a
quasi-periodic field
on $\rS^2$ and hence, via (\peq{nequal}), a field on $\rS^2/\Ga$ is
constructed by the projection
$$
\psi(\br)={1\over|\Ga|}\sum_{g\in\Ga}a(g)T_{g^{-1}}\widetilde\phi(\br).
\eql{proj}$$
In particular the projected (\ie adapted) modes are
$$
Y^{\Ga(l)}_{qm}(\br)={1\over\sqrt{|\Ga|}}\sum_{g\in\Ga}
a(g)\,\big(T_{g^{-1}}Y^{(l)}_{qm}
\big)(\br)=\sqrt{|\Ga|}\,Y^{(l)}_{qm'}(\br)P_{m'm}
\eql{pmodes}$$
where, using
(\peq{modact}),
$$
P_{m'm}={1\over|\Ga|}\sum_{g\in\Ga}a(g)\caD^{(l)}_{m'm}(g^{-1})
\eql{projm}$$
is a
hermitian projection matrix, $P^2=P$. Since the eigenvalues of $P$
are 0 and 1, this
shows that in a suitable, \ie diagonal, basis the projected eigenfunctions
are a subset of the unprojected ones, a general result and independent
of the magnetic field.

Depending on whether $q$ and $l$ are integral or half odd-integral
together, the representations $a$ and $\caD$ of the doubled
elements, $E\ga$,  have the same or opposite sign as those of
$\ga$ and we can see explicitly that it is adequate to restrict the
sum in (\peq{projm}) to $\Ga$, as we have done.

Let us pursue this a little further to make
sure everything is satisfactory. It is much more elegant to express
everything in abstract form but we will retain the coordinate
representation. The discussion is a textbook one in applied group theory.

From (\peq{pmodes}, using periodicity,  one easily derives the integral
$$
\int_\mod Y^{\Ga(l)*}_{qm}(\br) Y^{\Ga(l)}_{qm'}(\br)
d\br=P_{mm'}.
$$

Diagonalising $A$, $A=UDU^{-1}$, introduces the linear
combinations $ Y^{(\Ga)}U$, which, from (\peq{pmodes}) equals
$\sqrt{|\Ga|}\,YUD$ showing that certain linear combinations of the modes on
$\rS^2$ vanish and others do not, after adaptation which takes on the nature
of a filtering process. We write for the nonzero combinations
$$
\caY^{(l)}_{q\al}=\sqrt{|\Ga|}\,Y^{(l)}_{qm}U_{m\al}
\eql{curlyy}$$
which form a complete, orthogonal set on $\mod$.

The factors of $\sqrt{|\Ga|}$ have been chosen to give suitably
normalised eigenfunctions on $\rS^2/\Ga$ so that
$$
K_{\rS^2/\Ga}\big(\br,\br';\tau\big)=\sum_{l=|q|}^\infty
e^{-\la_l\tau}\,\Tr\left[\caY_q^{(l)}(\br)\caY_q^{(l)\dag}(\br')\right].
\eql{pkernel7}$$

The matrix $U$ reduces the $l$-representation of $\Ga^\bullet$
into irreducible ones and, from (\peq{projm}), the range of
$\al$, \ie the degeneracy,
is the number of times the irreducible $a$-representation occurs
in this decomposition.

\section{\bf 5. Effect of fixed points}
The standard theory of coverings applies when the covering group
acts freely. This is not so in the present case and the result is
a singular space -- an orbifold. We can still maintain the
standard covering terminology, as we already have done when
referring to fundamental domains, in an obvious way by firstly removing the
fixed points, so that the standard theory applies, and then extending the
action of the covering group, $\Ga$, to these points.

We now investigate the implications of the periodicity condition
(\peq{period}) when extended to the fixed point set of $\Ga$. For any
$\ga\in\Ga$ there are two fixed points on the sphere, $\pm\br_\ga$. Going
back to the mathematical result (\peq{modrot}), we see that the monopole
modes on the full sphere satisfy the {\it identity}
$$
Y^{(l)}_{qm'}(\pm\br_\ga)=e^{iq\La_\ga(\pm\br_\ga)}\,
Y^{(l)}_{qm}(\pm\br_\ga)\,\caD^{(l)}_{mm'}(\ga),
\eql{modide}$$
where $\ga$ is the rotation about $\pm\br_\ga$, obviously.

Using the relation (\peq{curlyy}) we can determine the action of the magnetic rotation
operator $T_\ga$ on the $\caY$,
$$
T_\ga\caY(\br)=\sqrt{|\Ga|}\, (T_\ga Y)(\br)
U=e^{-iq\La_\ga(\br)}\sqrt{|\Ga|}\,Y(\ga^{-1}\br)
U=e^{-iq\La_\ga(\br)}\caY(\ga^{-1}\br)
$$
as expected. However we also have, from (\peq{pmodes}),
$$
T_\ga\caY(\br)=a(\ga)\caY(\br)
$$
and therefore
$$
e^{-iq\La_\ga(\br)}\caY(\ga^{-1}\br)=a(\ga)\caY(\br)
$$
and of course
$$
e^{-iq\La_\ga(\br)}\psi(\ga^{-1}\br)=a(\ga)\psi(\br).
$$

If there were no fixed points there would be no problem to
discuss. However if $\br=\br_\ga$ is fixed by $\ga$ we have the
consistency condition
$$\eqalign{
e^{-iq\La_\ga(\br_\ga)}\psi(\br_\ga)&=a(\ga)\psi(\br_\ga)\cr
e^{-iq\La_\ga(\br_\ga)}\caY(\br_\ga)&=a(\ga)\caY(\br_\ga)\cr}
\eql{fp}$$

This says that the phase is undetermined at the fixed points.
If the fixed point is removed there is no problem. We are then
effectively working arbitrarily close to the fixed point and all
the points $\ga\br$, $\ga\in\Ga$, belong to different images of
the fundamental domain. When the fixed point is put back and the
limit $\br\to\br_\ga$ taken, all these images coalesce and the
phase becomes indeterminate, unless something special happens
such  as $e^{-iq\La_\ga(\br_\ga)}$ equalling $a(\ga)$ or if the
wave function has a node at $\br_\ga$. We analyse this situation further.

Since $\Ga$
can be generated by a set of cyclic rotations, it is helpful firstly to
take the case when $\Ga$ is cyclic, $C_k$, about the $z$-axis.
Everything in (\peq{projm}) is then explicit. Let the generator of
$C_k$ be $\widehat\ga$, ($\widehat\ga^{\,\nu}={\bf1}$). Also let
the generator of the double group $\oZ^\bullet$ be $\ga^\bullet$ with
$(\ga^\bullet)^\nu=E$ and $(\ga^\bullet)^{2\nu}={\bf 1}$. $E$ is the
$2\pi$ doubling rotation.

The representations $a(g^\bullet)$ are
$$
a\big((\ga^\bullet)^p\big)=e^{2\pi i pr/\nu},\quad p=0,1,\ldots,2\nu-1,
\eql{cycla}$$
where, for integral $q$, the label  $r$, is in the range $0$ to $\nu-1$,
while for half-odd integral $q$, $r=(2s+1)/2$ with $0\le s\le\nu-1$.

$\caD^{(l)}_{mm'}(g^\bullet)$ is diagonal
$$
\caD^{(l)}_{mm'}\big((\ga^\bullet)^p\big)=e^{-2\pi i pm/\nu}\,\de_{mm'},
\quad -l\le m,m'\le l,
\eql{cyclrep}$$
and
$P_{mm'}$ is easily found,
$$
P_{mm'}={1\over\nu}{1-e^{2\pi i(m+r)}\over 1-e^{2\pi
i(m+r)/\nu}}\,\de_{mm'}.
\eql{Aproj}$$
This expression vanishes unless $m+r$ is $0$
mod $\nu$, in which case it equals unity making the filtering obvious.

It is comforting to check things by looking at the explicit
forms of some modes. For $q=0$, the modes $Y^{(l)}_m$ are ordinary
spherical harmonics. At the north pole only the $m=0$ component
survives which then forces $r$ to be zero and (\peq{fp}) is
trivially satisfied since $a(\ga)=1$.

One can proceed generally but let us use the modes exhibited in
Wu and Yang section 7. For $q=1/2$ we see that only the $m=-1/2$ survives at
the north pole making $r=1/2$ and now to check (\peq{fp}) we need the
expression for $\La$. For a rotation through $\om$ about the
$z$-axis, from (\peq{phase}) or (\peq{phase2}), $\La=-\om$, and again the
check works. For $q=1$, the nonzero mode corresponds to $r=1$.

In general one finds that
$r=q$ and looking at the $\phi$ dependence of the modes,
$Y^{(l)}_{mq}$, (\peq{mon:harm1}),
\ie $\exp\big(i(m+q)\phi\big)$, we see that the modes $Y^{\Ga(l)}$
are periodic as $\phi$ increases by $2\pi/\nu$ and so,
therefore, is the wave function.

It should be noted that if (\peq{phase2}) is used, the
azimuthal angle of the north pole changes by $\om$ even though it
is a fixed point. The string's location may be unchanged, but
a nonzero compensating gauge transformation is still required.

It is clear geometrically that the same results will hold if $C_k$
is cyclic about any axis so that it is consistent to set
$$
a(\ga)=e^{-iq\La_{\ga}(\br_\ga)},
\eql{choice}$$
for any $\Ga$. The consequence of replacing $\br_\ga$ by $-\br_\ga$ is
discussed in the next section.

The minimal choice (\peq{choice}) clearly corresponds to untwisted
fields in the sense that the more general form
$$
a(\ga)=e^{-iq\La_\ga(\br_\ga)}\,b(\ga),
\eql{choice2}$$
where $b(\ga)$ is some nontrivial representation of $\Ga^\bullet$,
encodes physics over and above the magnetic monopole, such
as Aharonov-Bohm fluxes through the fixed points. In this latter case,
(\peq{period}) implies that the wavefunction vanishes at the fixed points.
\begin{ignore}
To confirm that the choice (\peq{choice}) is consistent it is
necessary to check the group combination rule which requires
$$
\La_{\xi\eta}(\br_{\xi\eta})=\La_\xi(\br_\xi)+\La_\eta(\br_\eta)
$$
or
$$
\La_\xi(\br_{\xi\eta})+\La_\eta(\xi^{-1}\br_{\xi\eta})
=\La_\xi(\br_\xi)+\La_\eta(\br_\eta).
$$
\end{ignore}

\section {\bf 6. Charge quantisation on rotational domains.}
We now have expressions for the modes and heat-kernel on
${\rS}^2/\Ga$ in the rotational case and have tacitly assumed that
all integer values of $2q$ are allowed. However $2q$ need only
form a subset of the integers. A geometric, Dirac type argument
will firstly be used to calculate this subset.

Choose a fundamental domain $\caF$ which does not contain the
string. Let $L\subset \caF$ be a piecewise continuous loop in
$\caF$ and construct the parallel propagator,
$$
\caI[L]=e^{-ie\int_L A}=e^{-iq\Om},
\eql{phasec}$$
where $A$ is the
potential one-form, $\Om$ is the area within $L$. This follows
from the explicit form of the monopole potential. As $L$ shrinks
to a point in $\caF$, $\caI[L]$ clearly tends to unity as the area
vanishes. Now $L$ also describes a loop with (oriented) area
$-(|\caF|-\Om)$ in $\rS^2/\Ga$, since $\rS^2/\Ga$ is closed for
rotational $\Ga$ (except for the fixed points which we can ignore
for these purposes as a set of measure zero). Thus if we let $L$
expand to fill the boundary $\pa\caF\subset \rS^2$, the propagator
will also take the value 1. This requires that $q|\caF|$ be a
multiple of $2\pi$ which gives the possible values of the magnetic
charge as $$ q=|\Ga|{n\over2},\quad n\in\oZ, \eql{quant3}$$ the
expected value since the magnetic charge per closed domain,
$\overline q\equiv q/|\Ga|$, equals the Dirac value, $n/2$.

If the string passes through $\caF$ there is an extra phase $-4\pi
q$ in (\peq{phasec}) which makes no difference to the final
result.

For the cyclic group, $C_k$, according to (\peq{cycla}), and
(\peq{choice}),
$$ a({\ga^\bullet}^p)=e^{2\pi ipq/\nu}=(-1)^{np}
\eql{quant4}$$
so all the $a(\ga)$ are $\pm1$.

A further restriction occurs if $C_k$ is a subgroup of a point group
$\Ga$, for then $$ a(\ga^\bullet)=e^{i\pi|\Ga|n/k}, $$ but, by
observation, for a noncyclic point group, all $|\Ga|/k$ are even
and the $a(\ga)$ are unity. Furthermore the monopole charge $q$ is
integral and there are no spinor modes.

The quantisation (\peq{quant3}) or (\peq{choice}), will now be derived
in another
way. The consistency condition (\peq{fp}) must hold for both fixed
points of $\ga$. Geometry shows, if orientation effects are
taken correctly into account, that $\La_\ga(-\br_\ga)=-
\La_\ga(\br_\ga)$. This also follows from the mode transformation
(\peq{modide}) and the behaviour under parity (\peq{parity}), which
involves $q\to -q$.

The two definitions of $a(\ga)$ require $$
q={k_\ga\over2} n_\ga \eql{cyclcond}$$ for some integer $n_\ga$,
where $k_\ga$ is the order of the cyclic subgroup generated by
$\ga$. Since $\La_\ga(\br_\ga)=2\pi/n_\ga$, all the $a(\ga)$ are
$\pm1$.

If $\Ga$ is not cyclic, the minimum requirement of
(\peq{cyclcond}) is $q=(K/2)p$ where $p\in\oZ$ and $K$ is the LCM
of all the $n_\ga$. Looking at the character tables of the double point
groups it is easily checked that it is not possible to reconcile
the tabulated $\pm1$ representations $a(\ga^\bullet)$ with (\peq{cyclcond})
unless all the $a(\ga^\bullet)$ are unity, implying that $p$ is even or
$q=(2K/2)p'$ for $p'\in\oZ$. But $2K$ is nothing but the group
order $\Ga$ leading to (\peq{quant3}) for all $n$. QED.

\section{\bf 7. Integrated kernels and zeta functions}

Special interest attaches itself to the integrated
kernel, $$\eqalign{
K_\Ga(\tau)\equiv&\int_{\rS^2/\Ga}K_{\rS^2/\Ga}\big(\br,\br;\tau\big)\,d\br
=\sum_{g\in\Ga}a(g)\int_{\rS^2/\Ga}\big(T_{g^{-1}}K_{\rS^2}\big)
\big(\br,\br;\tau\big)\,d\br\cr &={1\over|\Ga|}\sum_{g\in\Ga}a(g)
\int_{\rS^2}\big(T_{g^{-1}}K_{\rS^2}\big)\big(\br,\br;\tau\big)\,d\br.\cr}
\eql{tkernel}$$
It is convenient to write everything in terms of
the covering $\rS^2$ quantities because we can use the mode form
(\peq{kernel}) in (\peq{tkernel}) to obtain $$
K_\Ga(\tau)=\sum_{l=|q|}^\infty \, d_\Ga(l)\,e^{-\la_l\tau}
\eql{trker}$$ with
$$
d_\Ga(l)={1\over|\Ga|}\sum_ga(g)\chi^{(l)}(g^{-1})\,,
\eql{degen1}$$
where we have used the orthogonality of the $\rS^2$ monopole modes
and $\chi^{(l)}(g)$ is the usual ${\rm S O}^\bullet(3)\sim{\rm SU}(2)$
character of $g$. In particular
$$
\int_{\rS^2}\Tr\big[\big(T_g{\bf Y}_{ql}\big)(\br){\bf
Y}^\dagger_{ql}(\br)\big]\,d\br=\chi_l(g).
\eql{trmods}$$

Algebraically we can see that the degeneracy, $d_\Ga(l)$, is the
number of times the irreducible representation `$a$' occurs in the
$l$-representation. Since degeneracies are real, $a(g)$ in (\peq{degen1})
and in the following equations, can be replaced by its real part,
$\Real a(g)$, although in the present case the minimal $a(g)=\pm1$.

(\peq{degen1}) shows that the summand in $d_\Ga(l)$,
is a class function so that we can make a convenient
geometrical decomposition of the traced kernel, as in our earlier work
[\pref{ChandD}].
The preimage sum, \ie the sum over $\Ga$, is firstly
replaced by a sum over conjugacy classes, $\{g\}$,
$$
d_\Ga(l)={2l+1\over|\Ga|}+{1\over|\Ga|}\sum_{\{g\}}a(g)\,|\{g\}|\,\chi^{(l)}
(g)
$$
with $g$ being a typical element in $\{g\}$.

We now recall that the elements of a class correspond to rotations
through one fixed angle
about a set of conjugate axes. For a given set of such axes, one
corresponding class can be considered to be the primitive class, all
others associated with these axes then being generated by this one. Thus
the sum over all classes can be rewritten as a sum over primitive classes
and powers of these. Let $k$ be the generic order of the rotation
associated with the generic primitive class $\{\widehat g\}$ so that
$\widehat g^k=id$. Then $|\{\widehat g\}|$ is just the number, $n_k$, of
conjugate $k$-fold axes and we can write
$$
d_\Ga(l)={2l+1\over|\Ga|}+{1\over|\Ga|}\sum_{\widehat g}
n_k\sum_{p=1}^{k-1}a^p(\widehat g)\,\chi^{(l)}(\widehat g^p)
\eql{degen3}$$
which is the same as when there is no magnetic field, apart from
the restriction $l\ge|q|$ and one sees again, \cf , that the cyclic
groups, $C_k$, form the basic building blocks. This can be
made explicit as follows. In the case that
$\Ga$ is just $C_k$, one has from (\peq{degen3})
$$
d_k(l)={1\over k}\sum_{p=0}^{k-1}
a^p(\widehat g)\,\chi^{(l)}(\widehat g^p)
={2l+1\over k}+{1\over k}
\sum_{p=1}^{k-1}a^p(\widehat g)\,\chi^{(l)}(\widehat g^p)
\eql{cycldegen}$$
so that, substituting back,
$$
d_\Ga(l)={d_1(l)\over|\Ga|}\big(1-\sum_{\widehat g}n_k\big)
+{1\over|\Ga|}\sum_{\widehat g}k n_k\,d_k(l),
\eql{degen4}$$
where $d_1(l)=2l+1$ is the full sphere degeneracy.

Having now the degeneracies and the eigenvalues, we can turn to
the explicit construction of the integrated heat-kernel
(\peq{trker}) and its Mellin transform, the \zf,
$$
\ze_\Ga(s)=\sum_l {d_\Ga(l)\over\la_l^s}.
 \eql{zeta1}$$

The eigenvalues are given by (\peq{eigenv}) and we face the old
problem of computing spherical spectral quantities. We will
approach this by firstly looking at a `linearised' system, one whose
eigenvalues are $l+1/2$. The corresponding heat-kernel, denoted
$\widetilde K_\Ga(\tau)$, can be considered as that for
the pseudo-operator $\big(H_{\rS^2}+1/4+q^2\big)^{1/2}$
and will allow us to find the \zf\ for the eigenvalues $(l+1/2)^2$
quickly and that for the $\la_l$ of (\peq{eigenv}), more elaborately.
The expressions are also of some statistical mechanical interest
if $\tau$ is interpreted as an inverse temperature.

From (\peq{trker}), we have the linearised, or square root, kernel
$$
\widetilde K_\Ga(\tau)=\sum_{l=|q|}^\infty d_\Ga(l)e^{-(l+1/2)\tau}
\eql{linker}
$$
with degeneracies given by (\peq{degen1}).

In accordance with (\peq{degen4}), it is sufficient to consider the
cyclic group case, $K_k(\tau)$ and $\ze_k(s)$.

Since the cyclic axis is immaterial, we choose it to lie along the
$z$-axis and could use the expressions in the previous section since
$d_k(l)$ is $\Tr {\bf P}$, where ${\bf P}$ is given by
(\peq{Aproj}) with $r=q$. Therefore
$$
\widetilde K_k(\tau)={1\over k}\sum_{l=|q|}^\infty \sum_{m=-l}^l
e^{-(l+1/2)\tau}\,{1-e^{2\pi i(m+q)}\over 1-e^{2\pi i(m+q)/k}}.
\eql{linker2}
$$
The summations can be relabelled using (\peq{mon:harm2}) and performed, but
it is perhaps more elegant to substitute (\peq{degen1}) into (\peq{trker})
and do the $l$-summation first, as in [\pref{ChandD}].

Remembering the charge quantisation condition, (\peq{quant4}), the
degeneracies are
$$
d_k(l,q)={1\over k}\sum_{p=0}^{k-1}\cos(\pi np){\sin\big((2l+1)\pi
p/k)\over \sin(\pi p/k\big)},\quad q=nk/2,
\eql{qdegen}$$
where $n$ is even or odd. If $n$ is even, the degeneracies are
identical to the case when $q=0$, $d_k(l,q)=d_k(l,0)$.

We now introduce the generating function
$$
h_k(\si,q)=\sum_{l=|q|}^\infty d_k(l,q)\,\si^l
\eql{genf2}$$
closely connected with the traced heat-kernel, (\peq{trker}), if
$\si=e^{-\tau}$. For $n$ even the only effect of the monopole is to
make the series start at $l=|q|$.
However it is better to continue with the summations. We have, for both
integral and half  odd-integral $q$,
$$\eqalign{
h_k(\si,q)
&={1\over k}\sum_{l=|q|}^\infty\sum_{p=0}^{k-1}\cos(\pi np){\sin(2l+1)\pi p/k)
\over \sin(\pi p/k)}\,\si^l\cr
&={1\over k}\sum_{p=0}^{k-1}\cos(\pi np)
\sum_{l=|q|}^\infty{\sin(2l+1)\pi p/k)
\over \sin(\pi p/k)}\,\si^l\cr
&=\si^q{(1+\si+2q(1-\si))\over k(1-\si)^2}+
{1\over k}\sum_{p=1}^{k-1}\cos(\pi np)
\sum_{l=|q|}^\infty{\sin(2l+1)\pi p/k)
\over \sin(\pi p/k)}\,\si^l\cr
&=\si^q\bigg({1\over1-\si}{1+\si^k\over1-\si^k}+{2q\over k(1-\si)}\bigg)\cr}
\eql{genf3}$$
and so the corresponding heat-kernel is related in the simple
way
$$
\widetilde K_k^q(\tau)=e^{-q\tau}\bigg(\widetilde
K_k(\tau)+{2q\over k}\widetilde K_\infty(\tau)\bigg)
\eql{lincycl}$$
to the monopole-less ($q=0$) expression, [\pref{ChandD}],
$$
\widetilde K_k(\tau)={\coth(k\tau/2)\over2\sinh(\tau/2)}.
$$

The decomposition into conjugacy classes, (\peq{degen4}),
shows that this relation will follow through for all groups $\Ga$,
$$
\widetilde K_\Ga^q(\tau)=e^{-q\tau}\bigg(\widetilde K_\Ga(\tau)+
2\overline q\widetilde K_\infty(\tau)\bigg),
\eql{lingen}$$
where $\overline q$ is the charge per domain and $\widetilde K_\Ga(\tau)$
has been determined in [\pref{ChandD}],
$$
\widetilde
K_\Ga(\tau)={\cosh(d_0\tau/2)\over2\sinh(d_1\tau/2)\sinh(d_2\tau/2)}.
\eql{lkergen}$$
Here $d_0,d_1$ and $d_2$ are integer invariants associated with the
reflection group having $\Ga$ as its rotation subgroup.

The zeta function corresponding to the linear heat kernel is easily
determined as the Mellin transform of (\peq{lkergen}). In fact we shall find
it more useful to consider the slightly more general zeta function defined
by,
$$\eqalign{
\zeta^q_\Gamma(s,a)=&  {1\over\Gamma(s)}\int_0^\infty\tau^{s-1}
e^{\tau/2-a\tau}\widetilde K^q_\Gamma(\tau)d\tau\cr
\,&=\zeta_2(s,a+q|d_1,d_2)+\zeta_2(s,a+q+d_0|d_1,d_2)+
\bar{q}\zeta_H(s,a+q),\cr}
\eql{genzeta}$$
where $\zeta_H$ is the Hurwitz zeta function, and
$\zeta_2$ is the two-dimensional Barnes zeta function defined for $s>2$
by, [\pref{Barnesa}],
$$
\zeta_2(s,a|d_1,d_2)=\sum_{n_1,n_2=0}^\infty{1\over\big(a+n_1 d_1+n_2
d_2\big)^s}.
\eql{barnes}$$
$\ze_H$ is actually a one-dimensional Barnes \zf.

To conclude this section we shall present some properties of the Barnes
zeta function needed later.

The function $\zeta_2(s,a|d_1,d_2)$ has simple poles at $s=1,2$ whose
residues can be written in terms of generalised Bernoulli polynomials
$$
\Res_{s\to r}\zeta_2(s,a|d_1,d_2)=
\frac{(-1)^r}{d_1 d_2} B^{(2)}_{2-r}(a|d_1,d_2)
\eql{barnesres}$$
for $r=1,2$. Here we have used the more standard notation as in
Erdelyi [\pref{Erdelyi}]. The values of the Barnes \zf\ at negative
integers are also given in terms of generalised Bernoulli polynomials. For
$n\in\oZ^+$ we have,
$$
\zeta_2(-n,a|d_1,d_2)=\frac{1}{(n+1)(n+2)d_1 d_2}
B^{(2)}_{2+n}(a|d_1,d_2).
\eql{barnesneg}$$
The explicit generalised Bernoulli polynomials required in this paper are,
$$\eqalign{
B^{(2)}_0(a|d_1,d_2) = & 1 \cr
B^{(2)}_1(a|d_1,d_2) = & a-\half(d_1+d_2) \cr
B^{(2)}_2(a|d_1,d_2) = & a^2-(d_1+d_2)a+\frac{1}{6}(d_1^2+3d_1 d_2+d_2^2)
\cr
B^{(2)}_3(a|d_1,d_2) = & a^3-\frac{3}{2}(d_1+d_2)a^2+
\half(d_1^2+3d_1 d_2+d_2^2)a-\frac{1}{4}(d_1^2 d_2+d_1 d_2^2).\cr}
\eql{genbern}$$
Finally we present a useful Bernoulli identity which will
be used to simplify some expressions later on,
$$
B^{(2)}_n(d_1+d_2-a|d_1,d_2)=(-1)^n B^{(2)}_n(a|d_1,d_2).
\eql{bernid}$$

\section {\bf 8. Extension to reflection groups}
Before showing how to deal with the eigenvalues (\peq{eigenv}), we
extend the analysis to orbifolds $\rS^2/\Ga'$ where
$\Ga'$ is a finite reflection group -- the complete symmetry group of
a regular solid as outlined in section 1. The domain of interest
is a M\"obius triangle on $\rS^2$.

As shown in section 3, under
reflection, the magnetic charge, $q$, changes sign and the projection
has to take this into account. The rotational projection is given by
(\peq{pkernel}) and (\peq{pmodes}) and all that is necessary is to
combine this with the group decomposition (\peq{gdecomp}). We start by
writing down the projected modes
$$
W^{(l)}_{qm}(\br)=Y^{\Ga(l)}_{qm}(\br)+a(\si)T_{\si} Y^{\Ga(l)}_{-qm}
(\br),
\eql{reflmods}$$
where $\si$ is a reflection, say in one of the symmetry planes, so that
$a(\si)=\pm1$. We can choose either sign.

Note the extended periodicity condition
$$
T_{\ga'}W^{(l)}_{qm}(\br)=a(\ga)a(\si)W^{(l)}_{-qm}(\br),\quad
\ga'=\ga\si.
$$
which shows that the wavefunction on the domain $\ga'\caF$ has monopole
charge $-q$ if that on $\caF$ has charge $q$.

The group decomposition of $\Ga'$ gives
$$\eqalign{
\rS^2&=\bigg(\bigcup_{\ga\in \Ga}\overline{\ga \caF}\bigg)\cup
\bigg(\bigcup_{\ga'\in \Ga'_1}\overline{\ga'\caF}\bigg)\cr
&=\bigcup_{\ga\in \Ga}\overline{\ga({\bf1}+\si)\caF},\cr}
$$
and we see that the only possible theory using images requires that
adjacent domains on the sphere have opposite numerical monopole charge.
The charge is $q$ on $\ga\caF$ and $-q$ on $\ga'\caF$ for all $\ga\in\Ga$,
$\ga'\in\Ga'_1$.

For the theory to be consistent, we must at least show that the fundamental
domains with charge $q$ and $-q$ define equivalent physical theories.
There are two points. The first is that the sign of the charge is
arbitrary,
being essentially a matter of definition for the observer. The second
point is that any physically significant quantities will depend only on
$F_{\mu\nu}F^{\mu\nu}\sim q^2$. Our theory therefore has the
possibility of being consistent and we now derive the values of $q$
for which it {\it is} consistent.

Across the boundaries of the fundamental domains we have ${\bf B\to-\bf B}$
and ${\bf A\to-A}$. For consistency we should define these vector
quantities to
vanish on the reflecting boundaries. Consider now the parallel propagator,
(\peq{phasec}), where the loop $L\in\caF$. Since we have defined
${\bf A}$ to vanish
on $\pa\caF$ we have $\caI[\pa\caF]=1$. Thus for an arbitrary loop
approaching the boundary we require trivial parallel transport just as in the
pure rotational case. This gives the {\it possible} values of $q$ as
$$
q=|\Ga'|{n'\over2}=|\Ga|n',\quad n'\in\oZ,
\eql{quant2}$$
so $q$ is integral and there are no spinor modes.

The existence of fixed points imposes certain conditions. The
situation is more restricting than in the pure rotational case
because the fixed points form a continuous set, the boundary of
our domain, $\caF$. Let
$\si$ be a reflection ($\si^2={\bf1}$). The fixed point set, $\caP$,
is the intersection of the reflecting plane with the $\rS^2$ \ie a
great circle. Extending the periodicity condition (\peq{period})
to $\Ga'$ we see that $\psi$ would have to satisfy
$$
e^{-iq\Om_\si(\br)+iq\pi}\psi(\br)=a(\si)\psi(\br),\quad\forall\,
\br\in\caP.
\eql{fprefl}$$
Since $\Om_\si(\br)$ is not constant on $\caP$, this equation cannot be
satisfied on all of $\caP$ unless one of $q=0$, $\psi|_\caP=0$,
${\bf n }\perp \caP$ or
${\bf n}\in \caP$ is true. Applying this argument to the two, or three,
reflection generators removes the last two possibilities and we are
left with either $q=0$ or $\psi|_\caP=0$ (or both).

The fact that the magnetic charge, $q$, has to be of opposite sign
on adjacent domains under the action of $\Ga'$ for the image method to
work indicates, crudely, that $q$ is zero on the boundary and
suggests that our construction satisfies the restrictions following from
(\peq{fprefl}).

Developing this nonrigorous analysis, since the magnetic field $\bf B$
vanishes on the reflecting boundaries we must insist that the monopole
modes take the value $W^{(l)}_{0m}$ on the boundaries. Thus the consistency
equation (\peq{fprefl}) is satisfied. Let $\ga$ be an arbitary rotation
 in $\Ga$.
The rotational consistency (\peq{fp}) is satisfied for all $\ga$ since
$q$ is an integer multiple of $|\Ga|$ and, as already stated, $a(\ga)=1,
\forall \ga$. It would thus appear that all the values of $q$ in
(\peq{quant2}) produce a consistent theory.

For clarity we restate that the mode labelled by $qlm$ takes the
values
$$
\eqalign{W^{(l)}_{qm}\quad &{\rm on}\quad\Ga\caF\cr
W^{(l)}_{-qm}\quad &{\rm on}\quad\Ga_1'\caF\cr
W^{(l)}_{0m}\quad &{\rm on}\quad\pa(\Ga\caF)\sim-\pa(\Ga_1'\caF).\cr}
\eql{modsum}$$

The modes (\peq{reflmods}) can be used to define a heat-kernel
analogous to the rotational case.  We skip straight to the linear
heat-kernel which may be written
$$
\widetilde K_{\Gamma'}^q(\tau)=\int_{S^2/\Gamma'}\sum_{l=q}^\infty e^{-(l+1/2)\tau}
\Tr\big[{\bf W}^{(l)}_{q}(\br ){\bf W}^{(l)\dagger}_{q}(\br)\big]d\br.
$$

Here ${\bf W}^{(l)}_{q}$ is the vector of solutions $W^{(l)}_{qm}$. We can
extend this
integral to all $\rS^2$ just as we did in equation (\peq{tkernel}),
in the rotational case (this is not entirely trivial but it is
possible using the invariance of the theory under $q\to -q$),
$$
\widetilde K_{\Gamma'}^q(\tau)=\frac{1}{|\Gamma'|}\int_{S^2}\sum_{l=q}^\infty
e^{-(l+1/2)\tau} \Tr\big[{\bf W}^{(l)}_{q}(\br )
{\bf W}^{(l)\dagger}_{q}({\br})\big]d\br.
$$
The next step is to the use the explicit rotational modes
(\peq{pmodes}) or (\peq{curlyy})
(with $a(g)=1$), equations (\peq{modact}) and (\peq{reflact}), and the
invariance of the heat kernel under charge reversal. The result is,
$$
\widetilde K_{\Gamma'}^q(\tau)=\half \widetilde K^q_\Gamma(\tau)+
\frac{a(\si)}{|\Gamma'|}\sum_{\gamma'\in\Gamma_1}\int_{S^2}\sum_{l=q}^\infty
e^{-(l+1/2)\tau}\Tr\big[(T_{\gamma'}{\bf Y}^{(l)}_{-q})(\br)
{\bf Y}^{(l)\dagger}_q(\br)\big]d\br.
\eql{reflker}$$

This equation actually represents two heat kernels for the cases
$a(\si)=\pm 1$. For
$a(\si)=-1$ we shall use the term `Dirichlet' and write
$K^q_D(\tau)$. For $a(\si)=+1$ we use the term `Neumann' and write
$K^q_N(\tau)$. These names are used in analogy to the case with no monopole
field, $q=0$. In this case $W^{(l)}_{0\,m}$ is, by construction, a solution
on the whole sphere (see (\peq{modsum})) and must satisfy
$T_{\gamma'}W^{(l)}_{0\,m}=a(\si)W^{(l)}_{0\,m}$ everywhere. On the
reflecting boundaries, $a(\si)=-1$ requires $W^{(l)}_{0\,m}$ to vanish
\ie Dirichlet boundary conditions, and $a(\si)=+1$ requires the normal
derivative of $W^{(l)}_{0\,m}$ to vanish, \ie Neumann boundary conditions.

We now turn to some specific calculations of the heat kernels. Using
(\peq{trmods}) we can write the second term in (\peq{reflker}) as,
$$
\frac{a(\si)}{|\Gamma'|}\int_{S^2}\sum_{l=q}^\infty\sum_{\gamma'\in\Gamma_1}
\chi_l(\gamma') e^{-(l+1/2)\tau}.
\eql{ttrefl2}$$
Our first calculation is for the reflection group $\Gamma'$ with rotational
subgroup $C_k$. In this case $\Gamma'_1$ consists
of $k$ reflection planes with a common invariant axis, and with angle
$2\pi/k$
between adjacent planes. If we take one plane to be the $z-x$ plane then we
can
write $\Gamma'_1=\{\Pi g_\pi\hat\gamma^p|p=0,1,2,\ldots,k-1\}$ where $\Pi$ is
parity, $g_\pi$ is a rotation by angle $\pi$ about the $y$ axis, and
$\hat\gamma$
is a rotation by angle $2\pi/k$ about the $z$-axis. Using the explicit
result
$$
\caD^{(l)}{}_{mm'}(g_\pi)=(-1)^{l-m}\delta^m_{-m'}
\eql{piyrot}$$
and equations (\peq{modpar}), (\peq{cyclrep}) we find $\chi_l(\gamma')=1$
for all $\gamma'\in\Gamma'_1$ and all $l$. Thus (\peq{ttrefl2}) is trivial to
calculate in this case, and from (\peq{reflker}), (\peq{lincycl}) we find
$$
\widetilde K^q_D(\tau)=e^{-q\tau}{e^{-k\tau/2}\over 4\sinh(\tau/2)
\sinh(k\tau/2)}+\frac{q}{2k}{e^{-q\tau}\over\sinh(\tau/2)}
\eql{cyclrefld}$$
$$
\widetilde K^q_N(\tau)=e^{-q\tau}{e^{k\tau/2}\over 4\sinh(\tau/2)
\sinh(k\tau/2)}+\frac{q}{2k}{e^{-q\tau}\over\sinh(\tau/2)}.
\eql{cyclrefln}$$
The first term in each of these expressions is simply $\exp(-q\tau)$ times
the monopole-less heat kernel [\pref{ChandD}]. Notice that the extra monopole
contribution is the same in both heat kernels.

We can also calculate the heat kernels for the group $\Gamma'$ with
dihedral rotational subgroup $D_k$. We take $\Gamma'_1$ as $\si D_k$
where $\si$ is a reflection in the $x-y$ plane, and $D_k$ is the
dihedral group with the $k$-fold cyclic group about the $z$ axis, and
a rotation by $\pi$ about the $y$ axis. Using (\peq{cyclrep}) and
(\peq{piyrot}), we find $\chi_l(\gamma')=1$ for elements
$\gamma'\in\Gamma'_1$ involving the rotation by angle $\pi$ about $y$.
These elements are $|\Gamma|$ in number and contribute to (\peq{ttrefl2})
the expression
$$
\frac{a(\si)}{8}{e^{-q\tau}\over\sinh(\tau/2)}.
\eql{secbit}$$
The remaining elements can be written $\Pi g_\pi\hat\gamma^p$ where $g_\pi$
is a rotation
by $\pi$ about the $z$ axis, and $\Pi$, $\hat\gamma$ are as defined
above. Using (\peq{modpar}) and (\peq{cyclrep}) we find the transformation
matrix $(-1)^{l+m}\delta^m_{m'}\exp(-i 2\pi m p/k)$. This transformation is
non-trivial and we have to construct a sum similar to (\peq{linker2}).

For technical variety we use the relations
(\peq{mon:harm2}) so that the sum can be written,
$$\eqalign{
\frac{a(\si)}{4} e^{-\tau/2}\sum_{u=0}^\infty\sum_{t=-\infty}^\infty
(-1)^{kt}& (-e)^{-\tau(u+(|tk+q|+|tk-q|)/2)}\cr
&=a(\si) e^{-q\tau}{\cosh(k\tau/2)\over
8\cosh(\tau/2)\sinh(k\tau/2)}.\cr}
\eql{dihedsum}$$
Adding together (\peq{secbit}) and (\peq{dihedsum}), and using
(\peq{lincycl}) gives the results from (\peq{reflker}),
$$
\widetilde K^q_D(\tau)=e^{-q\tau}{e^{-(k+1)\tau/2}\over 4\sinh(2\tau/2)
\sinh(k\tau/2)}+\frac{q}{4k}{e^{-q\tau}\over\sinh(\tau/2)}
\eql{dihedrefld}$$
$$
\widetilde K^q_N(\tau)=e^{-q\tau}{e^{(k+1)\tau/2}\over 4\sinh(2\tau/2)
\sinh(k\tau/2)}+\frac{q}{4k}{e^{-q\tau}\over\sinh(\tau/2)}.
\eql{dihedrefln}$$
Again the first terms are simply $\exp(-q\tau)$ times the $q=0$ case.

For all the heat kernels derived above the extra monopole contribution is
simply given by
$$
\frac{q}{|\Gamma'|}{e^{-q\tau}\over \sinh(\tau/2)}.
\eql{extra}$$
Since we may construct the heat kernel for an arbitrary reflection group from
the heat kernels calculated above, [\pref{Meyer}], the simplicity of
(\peq{extra}) leads us to conclude that in the general case,
$$
\widetilde K^q_D(\tau)=e^{-q\tau}{e^{-d_0\tau/2}\over 4\sinh(d_1\tau/2)
\sinh(d_2\tau/2)}+\bar q'{e^{-q\tau}\over\sinh(\tau/2)}
\eql{genrefld}$$

$$
\widetilde K^q_N(\tau)=e^{-q\tau}{e^{d_0\tau/2}\over 4\sinh(d_1\tau/2)
\sinh(d_2\tau/2)}+\bar q'{e^{-q\tau}\over\sinh(\tau/2)}.
\eql{genrefln}$$
These two equations are the culmination of this section.

The first terms in these heat kernels are just $\exp(-q\tau)$ times the $q=0$
expressions [\pref{ChandD}], and we have defined the monopole charge per
reflection domain
$$
\bar q'=\frac{q}{|\Gamma'|}.
$$
Equation (\peq{reflker}) predicts that the linear Dirichlet and Neumann heat
kernels should add up to give the rotational linear heat kernel
(\peq{lingen}), and this is seen to be true (in performing this sum we must
use the relation $\bar q=2\bar q'$ for the same value of $q$).

The linear zeta functions for $\widetilde K^q_D$ and $\widetilde K^q_N$
are calculated using equation
(\peq{genzeta}) with $\widetilde K_\Gamma^q(\tau)$ suitably replaced. In
terms of the Barnes zeta function we find,
$$
\zeta^q_D(s,a)=\zeta_2(s,a+q+d_0|d_1,d_2)+\bar q'\zeta_H(s,a+q)\,.
\eql{genzetad}$$
$$
\zeta^q_N(s,a)=\zeta_2(s,a+q|d_1,d_2)+\bar q'\zeta_H(s,a+q)
\eql{genzetan}$$
It should be remembered that although we may add these two zeta functions to
produce the rotational zeta function, the physical theories are completely
different due to the changing sign of the magnetic field. Thus the procedure
can only be regarded as a formal trick.
\section{\bf9. General zeta function and its derivative}

Equations (\peq{genzeta}), (\peq{genzetad}) and (\peq{genzetan}) give zeta
functions for the eigenvalues $(l+a)^2$ if $s$ is replaced with $2s$. In
general we need the zeta functions for the more general eigenvalues
$\lambda_l=(l+a)^2-\alpha^2$, with suitable constants $a$ and $\alpha$. For
example we may add curvature coupling and mass terms to the
Hamiltonian $H_{\rS^2}$. In this case the eigenvalue equation is,
$$
(H_{\rS^2}+\xi R+m^2)Y^{(l)}_{qm}=\lambda_l Y^{(l)}_{qm}
$$
and we find $a=1/2$, $\alpha^2=q^2+(1/4-2\xi)-m^2$ ($R=2$ on the unit
two-sphere). It is assumed that $\alpha^2$ is positive. To analyse these
general zeta functions we use similar methods to those found in
[\pref{Dow1}]. For brevity we shall just write $\zeta(s)$ to represent a
general zeta function. For $s>1$, we have the explicit mode sum
$$
\zeta(s)=\sum_{l=|q|}^\infty{d(l)\over[(l+a)^2-\alpha^2]^s}.
\eql{czeta}$$
The function $d(l)$ is the degeneracy of the eigenvalue $\lambda_l$ for some
linear zeta function (i.e.\ rotational, Dirichlet or Neumann). If we assume
that $|\alpha|<|q|+a$ we can perform a binomial expansion on the summand
which leads to a continuation of $\zeta(s)$ given by
$$
\zeta(s)=\sum_{r=0}^\infty\alpha^{2r}{\Gamma(s+r)\over r!\Gamma(s)}
\zeta^q(2s+2r,a).
\eql{zbinom}$$
In the the above equation $\zeta^q(s,a)$ is intended to be any one of the
rotational, Dirichlet, or Neumann zeta functions defined by equations
(\peq{genzeta}), (\peq{genzetad}) and (\peq{genzetan}) respectively.

In order to tie in with [\pref{Dow1}] we generalise to
the case where $\zeta^q(s)$ represents an arbitrary zeta function on a
$d$-dimensional space with simple poles (only) at $s=1,2,\ldots,d$. This is
the situation encountered in [\pref{Dow1}] where $\zeta^q(s)$ would just be a
$d$-dimensional Barnes function $\zeta_d(s,a|\bv d)$ (of course the
label $q$ is defunct in the general case). Near the poles we define,
$$
\zeta^q(s+r,a)={N_r\over s}+R_r+O(s),\quad s\to 0,
$$
for $r=1,2,\ldots,d$. For our three cases (with $d=2$) the residues $N_r$ for
$r=1,2$ can be calculated from the specific forms of the zeta functions and
(\peq{barnesres}).

The important fact is that
the series (\peq{zbinom}) reduces to a finite sum when $s$ is a negative
integer. Thus we concentrate on these values of $s$ and find for $n\in\oZ^+$,
$$
(-\alpha^2)^{-n}\zeta(-n)=\sum_{r=0}^n(-\alpha^2)^{-r}\comb{n}{r}
\zeta^q(-2r,a)+\half\sum_{r=1}^u\alpha^{2r}\frac{n!(r-1)!}{(r+n)!}
N_{2r}.
\eql{mon:zeta4}$$
The number $u$ in the above is defined as $[d/2]$ where $d$ is the dimension
of the space under consideration. The derivative of the zeta function at
$s=-n$ can also be calculated from (\peq{zbinom}),
$$\eqalign{
(-\alpha^2)^{-n}\zeta'(-n)  =& 2\sum_{r=0}^n\comb{n}{r}(-\alpha^2)^{-r}
\zeta^q{}'(-2r,a)-\cr
\sum_{r=0}^n&\comb{n}{r}(-\alpha^2)^{-r}
(\psi(n+1)-\psi(r+1))\zeta^q (-2r,a)- \cr
&\sum_{r=1}^u\alpha^{2r}\frac{n!(r-1)!}{(r+n)!}\left\{R_{2r}+\half N_{2r}
\left(\psi(r)-\psi(n+1)\right)\right\}+\cr
&\sum_{r=u+1}^\infty\alpha^{2r}\frac{n!(r-1)!}{(r+n)!}\zeta^q(2r,a),\cr}
\eql{mon:zeta5}$$
where $\psi(z)=\Gamma'(z)/\Gamma(z)$ is the logarithmic derivative of
the gamma function.

The problem now rests on the evaluation of the infinite sum on the
last line of this expression. We show that this sum can be written in finite
terms. It is expected that the sum will be finite since we
expect $\zeta'(-n)$ to be finite, and all other terms on the right hand
side are finite. A word of caution is that in singular situations,
there is the possibility that logarithmic terms, $log\tau$, may appear in the
asymptotic expansion of the heat-kernel. If this were so, more
care would have to be taken over the evaluation of the
determinants. However no such terms occur here.

Using the integral representation (\peq{genzeta}) extended to the
arbitrary zeta
function $\zeta^q(s,a)$, we may write the last line in (\peq{mon:zeta5})
as,
$$
2n!\sum_{n=u+1}^\infty\alpha^{2r}\frac{r!}{(r+n)!(2r)!}\int_0^\infty
\tau^{2r-1}e^{\tau/2-a\tau}\widetilde K^q(\tau)d\tau\,,
\eql{mon:zsum1}$$
where $\widetilde K^q(\tau)$ is the linear heat kernel associated with
$\zeta^q(s,a)$. Since (\peq{mon:zsum1}) is assumed finite we may take the sum inside the
integral. Thus our problem can be reduced to evaluating the sequence of sums,
$$
T_n(\tau)=n!\sum_{r=1}^\infty\frac{r!\tau^{2r}}{(2r)!(r+n)!}.
$$

Using the simple result $\sqrt{\pi}(2r)!=2^{2r}r!\Gamma(r+1/2)$ and changing
the summation variable to $r'=r+n$ gives the result
$$
T_n(\tau)=n!\sqrt{\pi}(\half\tau)^{1/2-n}I_{-n-1/2}(\tau)
-\sum_{r=0}^n\comb{n}{r}(2r)!(-\tau^2)^{-r}\,,
$$
where $I_\nu(x)$ is the modified Bessel function.

To find a closed form for $T_n(\tau)$ we employ a useful integral
representation given in reference [\pref{MOS1966}]. For $\nu>0$,
$$\eqalign{
\Gamma(\half+\nu)&I_{-\nu}(x)=\cr
&\frac{2}{\sqrt{\pi}}(\half x)^\nu\left[
\int_{-1}^1\! e^{-xt}(1-t^2)^{\nu-1/2}dt+\sin(\pi\nu)\int_1^\infty
\!e^{-xt}(t^2-1)^{\nu-1/2}dt\right].\cr}
$$
Setting $\nu=n+1/2$ in this expression we may expand the integrand factors
$(1-t^2)^n$ using the binomial theorem leaving simple exponential
integrals. After a little work we find,
$$
T_n(\tau)=\sum_{r=0}^n\comb{n}{r}(-1)^r\left\{(-\tau^2)^{-n}(2\tau)^r (2n-r)!
\half\left(e^\tau+(-1)^r e^{-\tau}\right)-(2r)!\tau^{-2r}\right\}.
\eql{mon:Tn}$$

Having found a suitable expression for $T_n$ we can now go back to
(\peq{mon:zsum1}) and write it in the new form,
$$
\int_0^\infty\left\{2T_n(\alpha\tau)-\sum_{r=1}^u
(\alpha\tau)^{2r}\frac{n!(r-1)!}{(r+n)!(2r-1)!}\right\}
\tau^{s-1} e^{\tau/2-a\tau}\widetilde K^q(\tau)d\tau.
 \eql{mon:zsum2}$$
We have introduced into the integral a regulator $\tau^s$ (the expression that
we want is given for the value $s=0$). The continuation variable $s$ has been
introduced so that we may evaluate the integrals of the individual terms in the
sum definition (\peq{mon:Tn}) of $T_n$, and of the sum subtracted from it,
{\it before} we perform the sums. We assume that $s$ is large enough so that
all the individual integrals are well defined. In fact this requires
$s>2n$. Performing the integrations leaves,
$$
(-\alpha^2)^{-n}\sum_{r=0}^n\comb{n}{r}(2\alpha)^r (2n-r)!
\Gamma(s+r-2n)\times \hspace{2in}
$$
$$
\hspace{2in}
\left\{\zeta^q(s+r-2n,a+\alpha)+(-1)^r
\zeta^q(s+r-2n,a-\alpha)\right\}-
$$
$$
2\sum_{r=0}^n(-\alpha^2)^{-r}\comb{n}{r}(2r)!\Gamma(s-2r)\zeta^q(s-2r,a)
-\sum_{r=1}^u\alpha^{2r}
\frac{n!(r-1)!\Gamma(s+2r)}{(r+n)!(2r-1)!}\zeta^q(s+2r,a)
\eql{mon:zsum3}$$
As $s\to 0$ all of these terms diverge, although taking all terms together we
must get a finite result \ie all poles must cancel as $s\to 0$. This
cancellation of the poles leads to the equation,
$$
(-\alpha^2)^{-n}\sum_{r=0}^n\comb{n}{r}(2\alpha)^r
\left\{\zeta^q(r-2n,a-\alpha)+(-1)^r
\zeta^q(r-2n,a+\alpha)\right\}-
$$
$$
2\sum_{r=0}^n(-\alpha^2)^{-r}\comb{n}{r}\zeta^q(-2r,a)
-\sum_{r=1}^u\alpha^{2r}
\frac{n!(r-1)!}{(r+n)!}N_{2r}=0.
\eql{mon:pcancel}$$

Comparing equations (\peq{mon:pcancel}) and (\peq{mon:zeta4}) we see that the
pole cancellation is precisely the statement,
$$
\zeta(-n)=\half\sum_{r=0}^n\comb{n}{r}(2\alpha)^r
\left\{\zeta^q(r-2n,a-\alpha)+(-1)^r
\zeta^q(r-2n,a+\alpha)\right\}.
\eql{mon:szeta1}$$
This expression contains equally terms with arguments $a+\alpha$ and
$a-\alpha$. We shall say that $\zeta(-n)$ is `symmetric'. It is a
generalisation to general $n$ of the symmetric expression for $n=0$ found
in [\pref{Dow1}]. The methods used in this reference do not produce a
suitable pole cancellation to give a symmetric result for $\zeta(-n)$.

The finite remainder part of (\peq{mon:zsum3}) as $s\to 0$ is given by the
expression,
$$
(-\alpha^2)^{-n}\sum_{r=0}^n\comb{n}{r}(2\alpha)^r
\left\{\zeta^q{}'(r-2n,a-\alpha)+(-1)^r
\zeta^q{}'(r-2n,a+\alpha)\right\}-
$$
$$
2\sum_{r=0}^n(-\alpha^2)^{-r}\comb{n}{r}\zeta^q{}'(-2r,a)
-\sum_{r=1}^u\alpha^{2r}
\frac{n!(r-1)!}{(r+n)!}R_{2r}+
$$
$$
(-\alpha^2)^{-n}\sum_{r=0}^n\comb{n}{r}(2\alpha)^r \psi(2n-r+1)
\left\{\zeta^q(r-2n,a-\alpha)+(-1)^r
\zeta^q(r-2n,a+\alpha)\right\}-
$$

$$
2\sum_{r=0}^n(-\alpha^2)^{-r}\comb{n}{r}\psi(2r+1)\zeta^q(-2r,a)
-\sum_{r=1}^u\alpha^{2r}\frac{n!(r-1)!}{(r+n)!}\psi(2r)N_{2r}.
$$
Inserting this expression into (\peq{mon:zeta5}), and adding zero in the form
of $\psi(2n+1)$ times (\peq{mon:pcancel}) gives
$$\eqalign{
\zeta'(-n) &= \sum_{r=0}^n\comb{n}{r}(2\alpha)^r \left\{
\zeta^q{}'(r-2n,a-\alpha)+(-1)^r
\zeta^q{}'(r-2n,a+\alpha)\right\}- \cr
\,& \sum_{r=1}^n\comb{n}{r}(2\alpha)^r \sigma_r\left\{
\zeta^q(r-2n,a-\alpha)+(-1)^r
\zeta^q(r-2n,a+\alpha)\right\}- \cr}
$$
$$
\sum_{r=0}^{n-1}(-\alpha^2)^{n-r}\comb{n}{r}(2\psi(2r+1)-\psi(r+1)+
\psi(n+1)-2\psi(2n+1))\zeta^q(-2r,a)-
$$
$$
(-1)^n\sum_{r=1}^u\alpha^{2(r+n)}\frac{n!(r-1)!}{2(r+n)!}
(2\psi(2r)-\psi(r)+\psi(n+1)-2\psi(2n+1))N_{2r}.
\eql{zedashn}$$
The quantities $\sigma_r$ in the above equation are defined as the sums,
$$
\sigma_r=\sum_{k=0}^{r-1}\frac{1}{2n-k}.
$$
Equation (\peq{zedashn}) is not symmetric in the sense that, unlike
(\peq{mon:szeta1}), it does not depend only on the quantities $a\pm\alpha$. If
we assume that the sum over the residues is a true feature of $\zeta'(-n)$, as
it is for $\zeta'(0)$ in [\pref{Dow1}], then we are still left with a sum
over $\zeta^q(-2r,a)$. We will now re-write this sum in a more
natural, \ie symmetric, form.

To this end we introduce the intermediate zeta function,
$\overline\zeta(s)$, on the space
$\overline\man=\oR^{2n}\times \rS^2/\Gamma$, which is given by
$$
\overline\zeta(s)={\Gamma(s-n)\over (4\pi)^n
\Gamma(s)}\zeta(s-n)\,.
\eql{mon:bzeta1}$$
Combining this with (\peq{zbinom}) gives an expansion for
$\overline\zeta(s)$,
$$
\overline\zeta(s)=\sum_{r=0}^\infty\alpha^{2r}{\Gamma(s+r)\over r!\Gamma(s)}
\overline\zeta^q(2s+2r,a)
\eql{mon:bzeta2}$$
where we have defined the new linear zeta function via
$$
\overline\zeta^q(s,a)={\Gamma(\half s-n)\over(4\pi)^n\Gamma(\half s)}
\zeta^q(s-2n,a)
 \eql{mon:bzeta3}$$
The dimension of $\overline\man$ is $\overline d=2n+d$ (remember, for our
monopole case
$d=2$). We see from (\peq{mon:bzeta3}) that $\overline\zeta^q(s)$ has poles
 at
$s=2,4,\ldots,2n$ and $s=2n+1,2n+2,\ldots,\overline d$. The residues are
given by
the formulae,
$$
\overline N_{2r}={2 (-1)^r\over(4\pi)^n
(r-1)!(n-r)!}\zeta^q(2r-2n,a)\;,\quad
r=1,2,\ldots,n
\eql{mon:bNr1}$$
$$
\overline N_{2n+r}={\Gamma(\half r)\over(4\pi)^n\Gamma(\half r+n)}N_r
\;,\quad\quad r=1,2,\ldots,d\,.
\eql{mon:bNr2}$$

The purpose of making these new definitions is the functional similarity
between (\peq{zbinom}) and (\peq{mon:bzeta2}). This implies that results
like (\peq{mon:szeta1}) and (\peq{zedashn}) should exist for
$\overline\zeta(s)$
in terms of $\overline\zeta^q(s)$. The important point is that we know that
$\zeta'(0)$ can be written as a symmetric part and a sum over the residues of
$\zeta^q(s,a)$, either from [\pref{Dow1}], or setting $n=0$ in
(\peq{zedashn}). Thus we might expect $\overline\zeta'(0)$ to consist of a
symmetric part and a sum over the residues of
$\overline\zeta^q(s,a)$. Differentiating (\peq{mon:bzeta1}) and setting $s=0$
gives,
$$
\overline\zeta'(0)={(-1)^n\over (4\pi)^n n!}\big(\zeta'(-n)+
(\psi(n+1)-\psi(1))\zeta(-n)\big)\,.
\eql{mon:bzeta4}$$
Now $\zeta(-n)$ is symmetric in terms of $\zeta^q(s,a\pm\alpha)$ and
extends easily to a symmetric form for $\overline\zeta^q(s,a)$ using
(\peq{mon:bzeta3}). Thus by our reasoning we expect $\zeta'(-n)$ to contain a
sum over the residues $\overline N_r$. This is exactly what we
find, and the final result, in this section, is the {\it
symmetrical} expression, [\pref{Cook}],
$$\eqalign{
\zeta'(-n) &= \sum_{r=0}^n\comb{n}{r}(2\alpha)^r \left\{
\zeta^q{}'(r-2n,a-\alpha)+(-1)^r
\zeta^q{}'(r-2n,a+\alpha)\right\}- \cr
\,& \sum_{r=1}^n\comb{n}{r}(2\alpha)^r \sigma_r\left\{
\zeta^q(r-2n,a-\alpha)+(-1)^r
\zeta^q(r-2n,a+\alpha)\right\}- \cr
\,&(-1)^n(4\pi)^n  n!\sum_{r=1}^{\overline u}\frac{\alpha^{2r}}{r}
\rho_r\overline N_{2r},\cr}
 \eql{mon:szeta3}$$
with the definitions $\overline u=[\overline d/2]$ and
$$\eqalign{
\rho_r = & \psi(2r-2n+1)-\half\psi(r-n+1)-(\psi(2n+1)-\half\psi(n+1))\cr
\,  = &\sum_{k=0}^{r-1}\frac{1}{2k+1}-\sum_{k=0}^{n-1}\frac{1}{2k+1}.\cr}
$$

The conclusion of these manipulations is that, despite the apparent
awkwardness of the binomial expansion, (\peq{zbinom}), to obtain the
required \zf,
the quantities that we want are given in finite terms, (\peq{mon:szeta1}),
(\peq{zedashn}), (\peq{mon:szeta3}), and involve only
relatively standard functions such as generalised Bernoulli polynomials
introduced via the Barnes \zf.
%\begin{ignore}
\section{\bf10. Vacuum energy calculations}
Simply as an example of the use of the preceding expressions, we
evaluate some vacuum (Casimir) energies on $\rS^2/\Ga'$.

Let $\zeta^q(s,a)$ represent one of the rotational, Dirichlet or Neumann
linear
zeta functions as in the previous section. Then $\zeta^q(s,a)$ can be
extended
to the odd-dimensional space-time $\oR\times \rS^2/\Gamma$ (or $\oR\times
\rS^2/\Gamma'$) by
defining a new zeta function $\zeta(s)$ given by (with $a=1/2$),
$$
\zeta(s)={\Gamma(s-1/2)\over
(4\pi)^{1/2}\Gamma(s)}\zeta^q(2s,1/2).
 \eql{mon:ve0}$$
This zeta function corresponds to the rather artificial case of a
conformally coupled field in three dimensions with mass $q^2$.
\footnote{For a scalar field conformally coupled in $N$ dimensions we have
$4\xi=(N-2)/(N-1)$. In fact we only require that
$\alpha^2=0$ or equivalently $m^2=q^2+(1/4-2\xi)$.}
The vacuum energy associated with this physical situation is defined by the
simple formula [\pref{DandK}],
$$
E=-\half\mu_r\left.\frac{\rd}{\rd s}\left(\frac{\mu}{\mu_r}\right)^{2s}
\zeta(s)\right|_{s=0}.
\eql{mon:ve1}$$
In this equation $\mu$ is an arbitrary mass scale and $\mu_r=1/r$ is the mass
scale associated with the sphere radius $r$ (which has the value
$r=1$). Equation (\peq{mon:ve1}) is in fact just half the logarithmic
determinant. Inserting equation
(\peq{mon:ve0}) into (\peq{mon:ve1}) gives the simpler expression
$$
E=\half\zeta^q(-1,1/2),
\eql{mon:ve2}$$
which is finite and independent of $\mu$ (we have set $r=1$ again).

The vacuum energy on the space $\oR\times\oR^{2n}\times S^2/\Gamma(')$ can
also be found and gives $E$ proportional to
$\zeta^q(-1-2n,\half)$. The calculation for general $n$ is entirely
equivalent
to the $n=0$ case, which we now calculate.

Using the definitions (\peq{genzetad}), (\peq{genzetad}) and equation
(\peq{barnesneg}), we find from (\peq{mon:ve2}) the
Dirichlet and Neumann vacuum energies,
$$
E_{\left\{ {D \atop  N} \right\}}=\pm\frac{d_0}{48|
\Gamma'|}(d_0^2-d_1^2-d_2^2)
+\frac{\overline q'}{24}(3d_0^2-d_1^2-d_2^2+\half\pm 6d_0q-2q^2).
$$
The constant ($q$ independent) terms are exactly the same as those calculated
in [\pref{ChandD}] for $q=0$, as required. Adding together the Dirichlet
and Neumann vacuum energies gives the rotational vacuum energy
$$
E_\Gamma=\frac{\overline q}{24}(3d_0^2-d_1^2-d_2^2-2q^2),
$$
where we have used the relation $\overline q=2\overline q'$ for
fixed $q$. This vacuum
energy necessarily vanishes for $q=0$, as proved in [\pref{ChandD}].
We now list the vacuum energies $E_{\left\{D\atop N\right\}}$ for all
possible reflection groups $\Gamma'$,
$$\eqalign{
O^*&:\quad\pm\frac{29}{256}+\frac{\overline q'}{48}(383\pm 108q-4q^2)\cr
Y^*&:\quad\pm\frac{89}{384}+\frac{\overline q'}{48}(1079\pm 90q-4q^2)\cr
O]T&:\quad\pm\frac{11}{192}+\frac{\overline q'}{48}(167\pm 36q-4q^2)\cr
D_n]C_n&:\quad\mp\frac{1}{96}+\frac{\overline q'}{48}(4n^2-1\pm 6nq-4q^2)\cr
D_n^*  (n{\rm\ even}),\quad  D_{2n}]D_n (n{\rm
\ odd})&:\quad
\pm\frac{(n+1)(2n-3)}{192n}
\frac{\overline q'}{48}(4(n+1)(n+2)-9\pm\cr
&\hspace{*****************}6(n+1)q-4q^2).\cr}
$$
Those for the corresponding rotational subgroups are obtained by
adding the D and N values.
\begin{ignore}
Rotation group & $E_\Gamma$ \\\hline
$O$ & $\frac{\overline q}{48}(383-4q^2)$ \\
$Y$ & $\frac{\overline q}{48}(1079-4q^2)$ \\
$T$ & $\frac{\overline q}{48}(167-4q^2)$ \\
$C_n$ & $\frac{\overline q}{48}(4n^2-1-4q^2)$ \\
$D_n$ & $\frac{\overline q}{48}(4(n+1)(n+2)-9-4q^2)$ \\\hline
\end{ignore}

\begin{ignore}
\begin{table}[!t]

\begin{center}
\begin{tabular}{|c|c|c|c|}\hline
Reflection group & $E_{\left\{D\atop N\right\}}$ \\\hline
$O^*$ & $\pm\frac{29}{256}+\frac{\overline q'}{48}(383\pm 108q-4q^2)$ \\
$Y^*$ & $\pm\frac{89}{384}+\frac{\overline q'}{48}(1079\pm 90q-4q^2)$ \\
$O]T$ & $\pm\frac{11}{192}+\frac{\overline q'}{48}(167\pm 36q-4q^2)$ \\
$D_n]C_n$ & $\mp\frac{1}{96}+\frac{\overline q'}{48}(4n^2-1\pm 6nq-4q^2)$
\\
$\left.\begin{array}{ll} D_n^* & (n{\rm\ even}) \\ D_{2n}]D_n & (n{\rm
\ odd})
\end{array}\right\}$ & $\pm\frac{(n+1)(2n-3)}{192n}+
\frac{\overline q'}{48}(4(n+1)(n+2)-9\pm 6(n+1)q-4q^2)$ \\\hline\hline
Rotation group & $E_\Gamma$ \\\hline
$O$ & $\frac{\overline q}{48}(383-4q^2)$ \\
$Y$ & $\frac{\overline q}{48}(1079-4q^2)$ \\
$T$ & $\frac{\overline q}{48}(167-4q^2)$ \\
$C_n$ & $\frac{\overline q}{48}(4n^2-1-4q^2)$ \\
$D_n$ & $\frac{\overline q}{48}(4(n+1)(n+2)-9-4q^2)$ \\\hline
\end{tabular}
\end{center}

\caption[Vacuum energy on $\oR\times S^2/\Gamma'$.]{Vacuum energy on
$\oR\times
S^2/\Gamma'$: This table contains the vacuum energies for all reflection
groups
in table~\ref{tab:groups}. Where there is a $\pm$ or $\mp$, the upper sign
corresponds to Dirichlet and the lower to Neumann boundary conditions.}
\label{tab:monenergy}

\end{table}
\end{ignore}

We now go on to calculate the vacuum energy on the even-dimensional space
$\oR^{2n}\times S^2/\Gamma(')$. Using equations (\peq{mon:bzeta1}),
(\peq{mon:bzeta4}) and (\peq{mon:ve1}) we find the expression,
$$
E={(-1)^{n+1}\mu_r\over 2(4\pi)^n n!}\left(\zeta'(-n)+\left[
\ln\left(\frac{\mu}{\mu_r}\right)^2+\psi(n+1)-\psi(1)\right]\zeta(-n)\right).
\eql{mon:ve3}$$
Here $\zeta(s)$ is the general zeta function defined via equations
(\peq{czeta}) and (\peq{zbinom}).

An infinite contribution has been (arbitrarily) dropped to arrive at
(\peq{mon:ve3}), the logarithmic term being a relic of this divergence.
Since $\zeta(0)$ is not zero for any of the three monopole theories, we
conclude from (\peq{mon:ve3}) that the vacuum energy is explicitly dependent
on the arbitrary scale $\mu$. However for simplicity we shall assume
$\mu=\mu_r=1$ for the rest of this section. Our concern in this paper is
not with realistic quantum field theory considerations.

Equations (\peq{mon:szeta1}),
(\peq{zedashn}) imply that the calculation for increasing $n$ merely
requires the evaluation of more and more zeta functions and their derivatives.
Thus we shall concentrate on the simplest case $n=0$ corresponding
to $\rS^2/\Gamma(')$ itself. This will also allow us to compare with the
results
for $q=0$ studied in [\pref{Dow1}].

We consider the case of a massless field with minimal coupling, that is
$m^2=0$
and $\xi=0$. Thus we have $a=1/2$ and $2\alpha=\sqrt{4q^2+1}$ (we shall still
write $\alpha$ when convenient). From the definitions (\peq{genzetad}),
(\peq{genzetan}), and equations (\peq{mon:szeta1}), (\peq{barnesneg}),
we find for Dirichlet and Neumann boundary conditions
$$
\zeta_{\left\{D\atop N\right\}}(0)=\frac{1}{12}+\frac{1}{6|\Gamma'|}(
d_0(d_0-1)+1\pm 6d_0 q+18q^2).
$$
These expressions reduce to those found earlier for
$q=0$ [\pref{Dow1}]. For the rotational case we simply add the Dirichlet
and Neumann results to give
$$
\zeta_\Gamma(0)=\frac{1}{6}+\frac{1}{6|\Gamma|}(d_0(d_0-1)+1+18q^2).
$$

To calculate the vacuum energy we have still to calculate
$\zeta'(0)$. Considering equation (\peq{zedashn}) with $n=0$ requires the
evaluation of the residue $N_2$. For both Dirichlet and Neumann zeta
functions we find from (\peq{barnesres}) and (\peq{genbern}) the value
$N_2=2/|\Gamma'|$. Using the derivative of the Hurwitz zeta function in
[\pref{Erdelyi}] then gives the zeta function derivatives
$$\eqalign{
\zeta'_D(0) = & \zeta'_2(0,\half+q-\alpha+d_0)+\zeta'_2(0,\half+q+\alpha+d_0)
+ \cr &
\overline q'\ln\left\{\Gamma(\half+q-\alpha)\Gamma(\half+q+\alpha)\right\}-
\overline q'\ln(2\pi)-\frac{1+4q^2}{2|\Gamma'|}\,,\cr}
\eql{mon:zNp0}$$
$$\eqalign{
\zeta'_N(0) = & \zeta'_2(0,\half+q-\alpha)+\zeta'_2(0,\half+q+\alpha)
+ \cr &
\overline q'\ln\left\{\Gamma(\half+q-\alpha)\Gamma(\half+q+\alpha)\right\}-
\overline q'\ln(2\pi)-\frac{1+4q^2}{2|\Gamma'|}\,.\cr}
\eql{mon:zDp0}$$
(The rotational zeta function derivative is just the sum of these two.) The
triangle inequality $|x|+|y|\geq\sqrt{x^2+y^2}$ ($x,y\in\oR$) implies that
$1/2+q\geq\alpha$. The equality is only met for $q=0$, and in this case we
have to remove the singularity in the first term
$\zeta'_2(0,1/2+q-\alpha|d_1,d_2)$ of the Dirichlet zeta function as
in~\pref{Dow1} (we shall do this later). For $q>0$ all terms in
(\peq{mon:zDp0}) and (\peq{mon:zNp0}) are well defined.

There is no known analytic form for the derivatives of the Barnes zeta
functions appearing in (\peq{mon:zDp0}), (\peq{mon:zNp0}) and so we have
to calculate them numerically. To do this we obviously need a
continuation of
the Barnes zeta function which is open to easy numerical computation. In
[\pref{Dow1}] several efficient continuations are presented
which are valid for positive integer values of $d_1$ and $d_2$.  However
as we
shall show in the next section, it is useful to have an expression which is
valid for all $d_1,d_2\in\oR^+$. We shall now derive such an expression.

Our starting point is the Plana sum formula which we display here
[\pref{Lindel}], $$\eqalign{ \sum_{n=a}^b f(n) =&
\half(f(a)+f(b))+\int_a^b f(t)dt+ \cr &
i\int_0^\infty{f(a+it)-f(a-it)-f(b+it)+f(b-it)\over e^{2\pi t}-1}dt\,.\cr}
\eql{mon:planar}$$
To be valid $f(t)$ must be an
analytic function in the region of the complex $t$~plane
$a\leq\Real(t)\leq b$, and the integrals must exist. Applying
(\peq{mon:planar}) twice to the sum definition (\peq{barnes})
of the Barnes zeta function gives immediately, for $s>2$,
$$\eqalign{ \zeta_2(s,a|d_1,d_2) =&\half
d_1^{-s}\zeta_H(s,\frac{a}{d_1})+
\frac{1}{d_2(s-1)}d_1^{-s}\zeta_H(s-1,\frac{a}{d_1})+ \cr \,&
i\int_0^\infty{dt\over e^{2\pi
t}-1}\bigg\{\half\left((a+id_2t)^{-s}- (a-id_2t)^{-s}\right)+ \cr
&\frac{1}{d_1(s-1)}\left((a+id_2t)^{1-s}-(a-id_2t)^{1-s}\right)\bigg\}+
\cr & \int_0^\infty{du\over e^{2\pi u}-1}\int_0^\infty{dt\over
e^{2\pi t}-1} \bigg\{ (a+id_1u-id_2t)^{-s}+ \cr &
(a-id_1u+id_2t)^{-s}-(a+id_1u+id_2t)^{-s}-(a-id_1u-id_2t)^{-s}\bigg\}
\;. \cr} \eql{mon:barnes3}$$ It is simple to verify that we are
meeting the conditions required for the validity of the sum
formula. In order to get rid of the single integrals in
(\peq{mon:barnes3}) we use the Plana sum definition of the Hurwitz
zeta function which is, from
(\peq{mon:planar})
$$ \zeta_H(s,a)=\half
a^{-s}+\frac{a^{1-s}}{s-1}+i\int_0^\infty {dt\over e^{2\pi
t}-1}\left\{(a+it)^{-s}-(a-it)^{-s}\right\}.
 \eql{mon:hzeta1}$$
The integral part of this expression is equivalent to the integrals
appearing in (\peq{mon:barnes3}).

To simplify the double integral in (\peq{mon:barnes3}) we first perform a
change of variables from $t,u$ to $d_1 u\pm d_2t$. Following this, we use
the easily proved formula,
$$
(x+iy)^s+(x-iy)^s=2\cos\big(s\tan^{-1}{y}/{x}\big),
$$
which is valid for $\Real(x)\geq 0$. After a little work we find the more
convenient form for the double integral,
$$
\frac{2}{d_1 d_2}\int_0^\infty dw\,{G(w)\over e^{2\pi w}-1}
\;{\cos\big(s\tan^{-1}(w/a)\big)\over (a^2+w^2)^{s/2}}.
\eql{mon:double}$$
All the non-trivial $d_1$, $d_2$ dependence has been absorbed into the
function $G(w)$ which is independent of $s$ and has the explicit form
$$\eqalign{
G(w)=&(e^{2\pi w}-1)\bigg\{
\int_0^w{dy\over(e^{\delta_1 y}-1)(e^{\delta_2(w-y)}-1)}- \cr &
\int_0^\infty{dy\over(e^{\delta_1 y}-1)(e^{\delta_2(w+y)}-1)}-
\int_0^\infty{dy\over(e^{\delta_1(w+y)}-1)(e^{\delta_2 y}-1)}\bigg\}\cr}
\eql{mon:G1}$$
where we have defined $\delta_i=2\pi/d_i$, $i=1,2$. This function is
symmetric
under the interchange of $d_1,d_2$ as one would expect. The factor
$(e^{2\pi w}
-1)$ has been included into the definition to ensure that $G(w)$ is finite as
$w\to 0$.

All the integrals in the definition of $G(w)$ are divergent at their lower
limits, and the first integral is also divergent at its upper limit. However
one can check that the combination is well defined. In fact by expanding the
integrands at their limits of integration we find that for small
$\epsilon>0$,
$$\eqalign{
G(w)=&(e^{2\pi w}-1)\bigg\{\int_\epsilon^{w-\epsilon}
{dy\over(e^{\delta_1 y}-1)(e^{\delta_2(w-y)}-1)}- \cr &
\int_\epsilon^\infty{dy\over(e^{\delta_1 y}-1)(e^{\delta_2(w+y)}-1)}-
\int_\epsilon^\infty{dy\over(e^{\delta_1(w+y)}-1)(e^{\delta_2 y}-1)}\bigg\}
+O(\epsilon)\,.\cr}
$$
Thus $G(w)$ is easy to calculate numerically, with the error being of order
$\epsilon$. We may also make $\epsilon$ the lower limit of the integration
over
$w$ in (\peq{mon:double}). Since the integrand with respect to $w$ is finite
at
the lower limit, the error incurred will still be $O(\epsilon)$.

The full expression for our continuation of the Barnes zeta function, after
dealing with both the single and double integrals in (\peq{mon:barnes3}), is
$$\eqalign{
\zeta_2(s,a|d_1,d_2) =& -\frac{1}{4}a^{-s}-\frac{a^{2-s}}{d_1 d_2(s-1)(s-2)}-
\frac{(d_1+d_2)a^{1-s}}{2d_1 d_2(s-1)}+ \cr
& \half d_1^{-s}\zeta_H(s,\frac{a}{d_1})+
\half d_2^{-s}\zeta_H(s,\frac{a}{d_2})+ \cr &
\frac{1}{d_1 d_2(s-1)}\left\{d_1^{2-s}\zeta_H(s-1,\frac{a}{d_1})+
d_2^{2-s}\zeta_H(s-1,\frac{a}{d_2})\right\}+ \cr &
\frac{2}{d_1 d_2}\int_0^\infty {G(w)dw\over e^{2\pi w}-1}
\;{\cos\big(s\tan^{-1}(w/a)\big)\over(a^2+w^2)^{s/2}}\cr}
\eql{mon:barnes4}$$
Although this formula was derived for $s>2$, it is actually a continuation to
all values of (complex) $s$ except at the points $s=1,2$ where there are
simple
poles. It is easy to check that these poles are correct in that their
residues
match those given in (\peq{barnesres}). From (\peq{mon:barnes4}) we can
calculate the derivative of the Barnes zeta function at $s=0$,
$$\eqalign{
\zeta'_2(0,a|d_1,d_2) =& -\frac{1}{4}\ln a-\half\left(1-\frac{a}{d_1}-
\frac{a}{d_2}\right)\ln(2\pi)+\half a\ln a\left(\frac{1}{d_1}+
\frac{1}{d_2}\right)- \cr
&\frac{a^2}{2d_1 d_2}\left(\frac{5}{2}+\ln a\right)+
\left(\half-\frac{a}{d_1}\right)\left(\half-\frac{a}{d_2}\right)
\ln\frac{a^2}{d_1 d_2}+ \cr
& \left(\half-\frac{a}{d_1}\right)\ln\Gamma(\frac{a}{d_2})+
\left(\half-\frac{a}{d_2}\right)\ln\Gamma(\frac{a}{d_1})+
\frac{d_1^2+d_2^2}{12d_1 d_2}+ \cr
\hspace{1in}\frac{1}{d_1 d_2}&\int_0^\infty {dw\over e^{2\pi w}-1}
\;\big(2G(w)-d_1^2w\ln(a^2+d_1^2w^2)-d_2^2w\ln(a^2+d_2^2w^2)\big)\cr}
\eql{mon:barnes5}$$
Here we have used (\peq{mon:hzeta1}) again to convert derivatives of the
Hurwitz zeta function $\zeta_H(s,a)$ at $s=-1$ into integrals suitable for
numerical evaluation.

The expression (\peq{mon:barnes5}) can be used directly in equation
(\peq{mon:zDp0}) and (\peq{mon:zNp0}). Figures 1 to
4 show plots of $E=-\zeta'(0)/2$ for small values of $\overline q$.
The lines on the graphs for $T$, $O$, and $Y$ are for labelling only. The
graphs for $C_k$ and $D_k$ are plotted for all $k$, although at the moment we
are only concerned with the integral values $k=1,2,\ldots$ marked with
crosses. On each graph the value $\overline q$ of the charge per rotational
domain is always twice the reflection value $\overline q'$ so as to give the
same value of $q$.
\newpage
\epsfxsize=350pt
\epsfbox{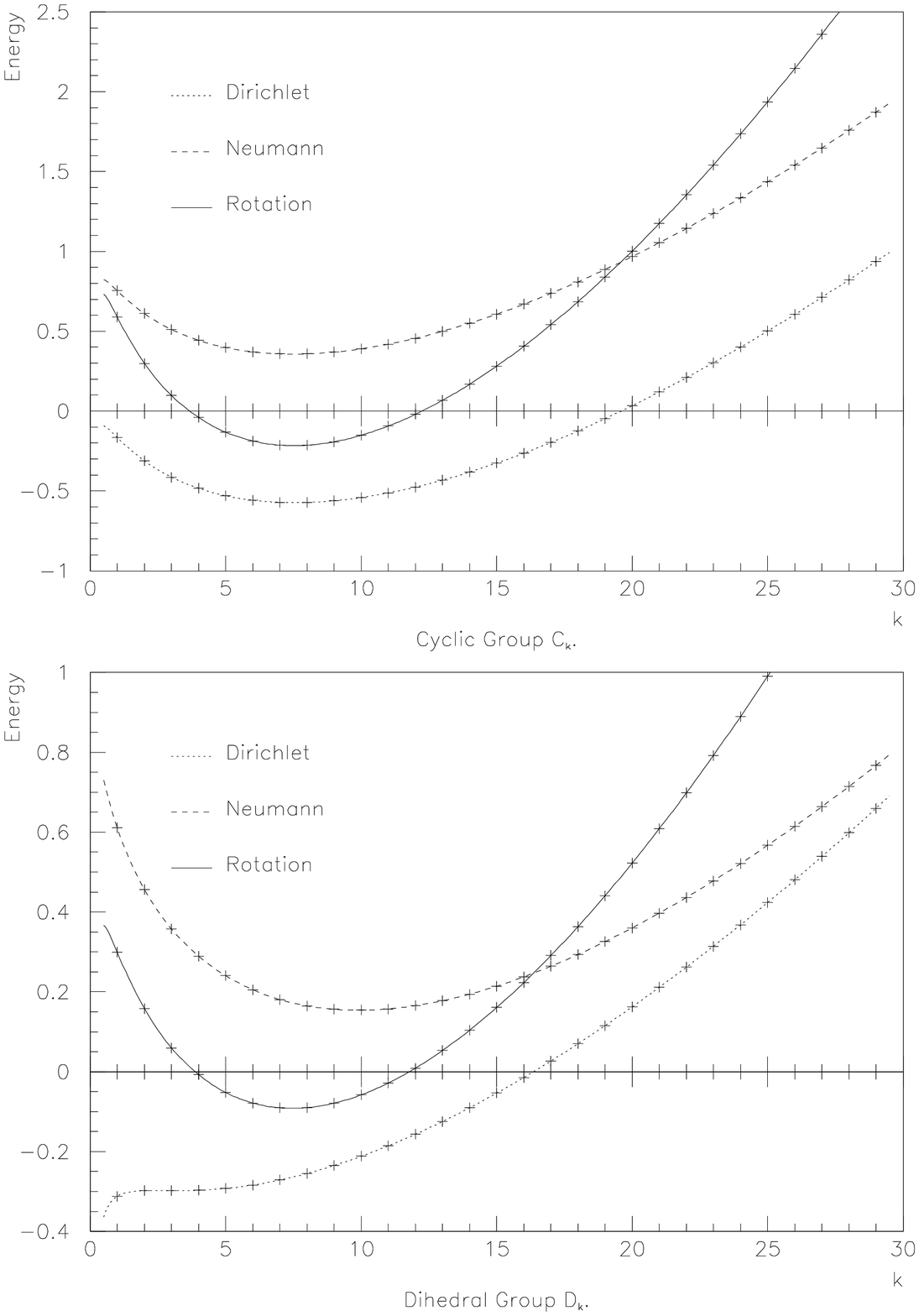}
Fig.1  Vacuum energies for $C_k$ and $D_k$ with $\overline q'=0$.
\newpage
\epsfxsize=350pt
\epsfbox{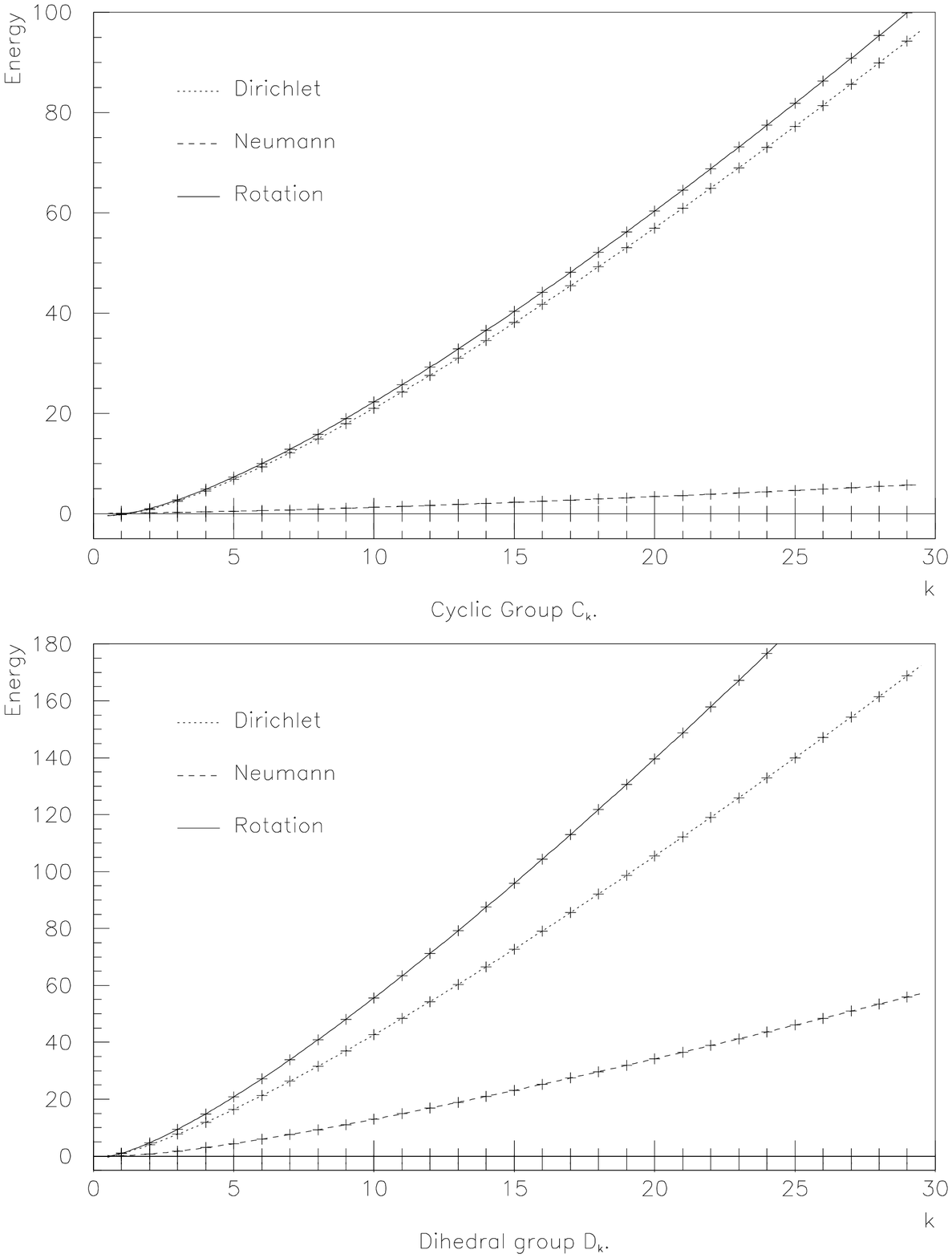}
Fig.2  Vacuum energies for $C_k$ and $D_k$ with $\overline q'=1/2$.
\newpage
\epsfxsize=350pt
\epsfbox{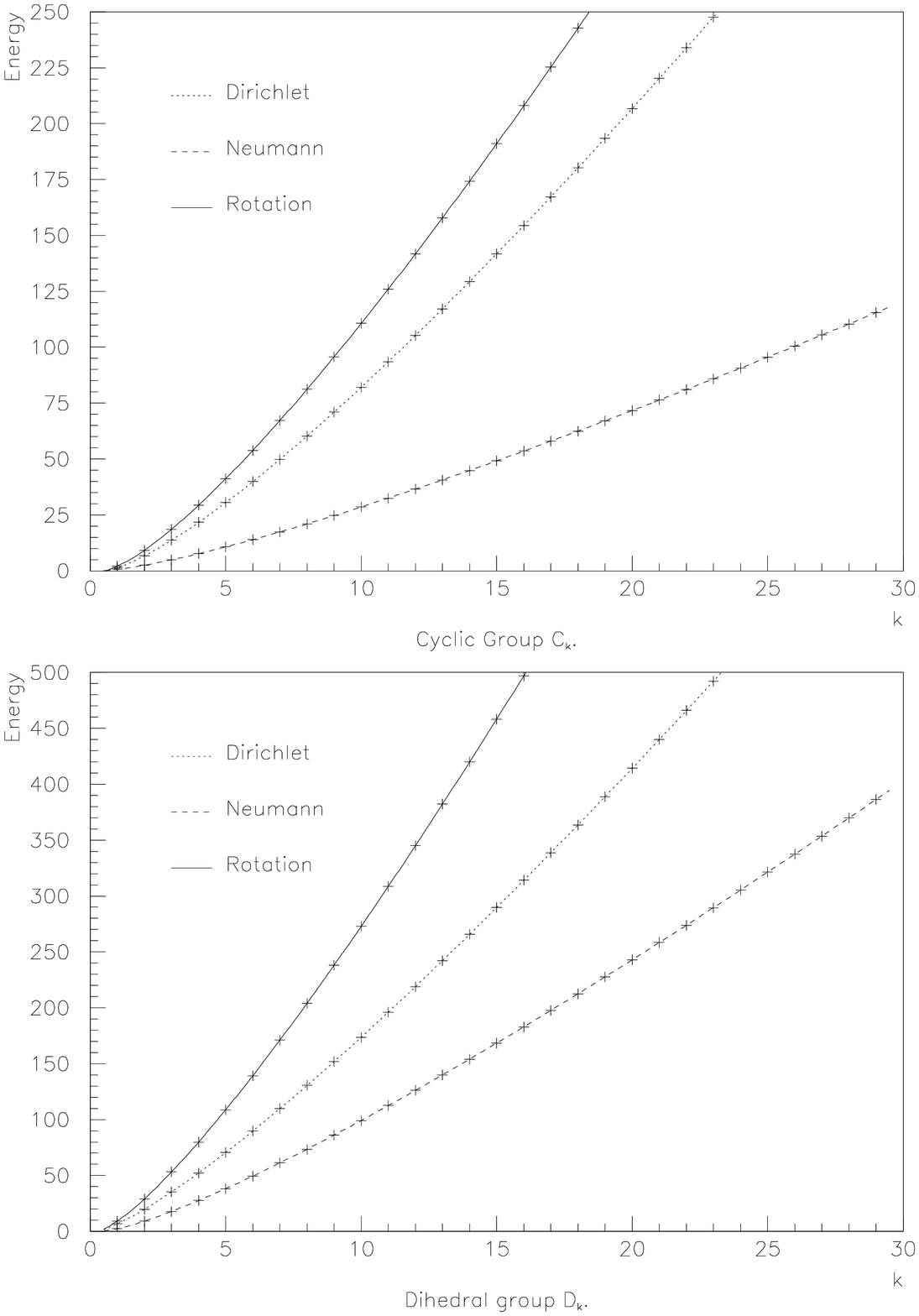}
Fig.3  Vacuum energies for $C_k$ and $D_k$ with $\overline q'=1$.
\newpage
\epsfxsize=350pt
\epsfbox{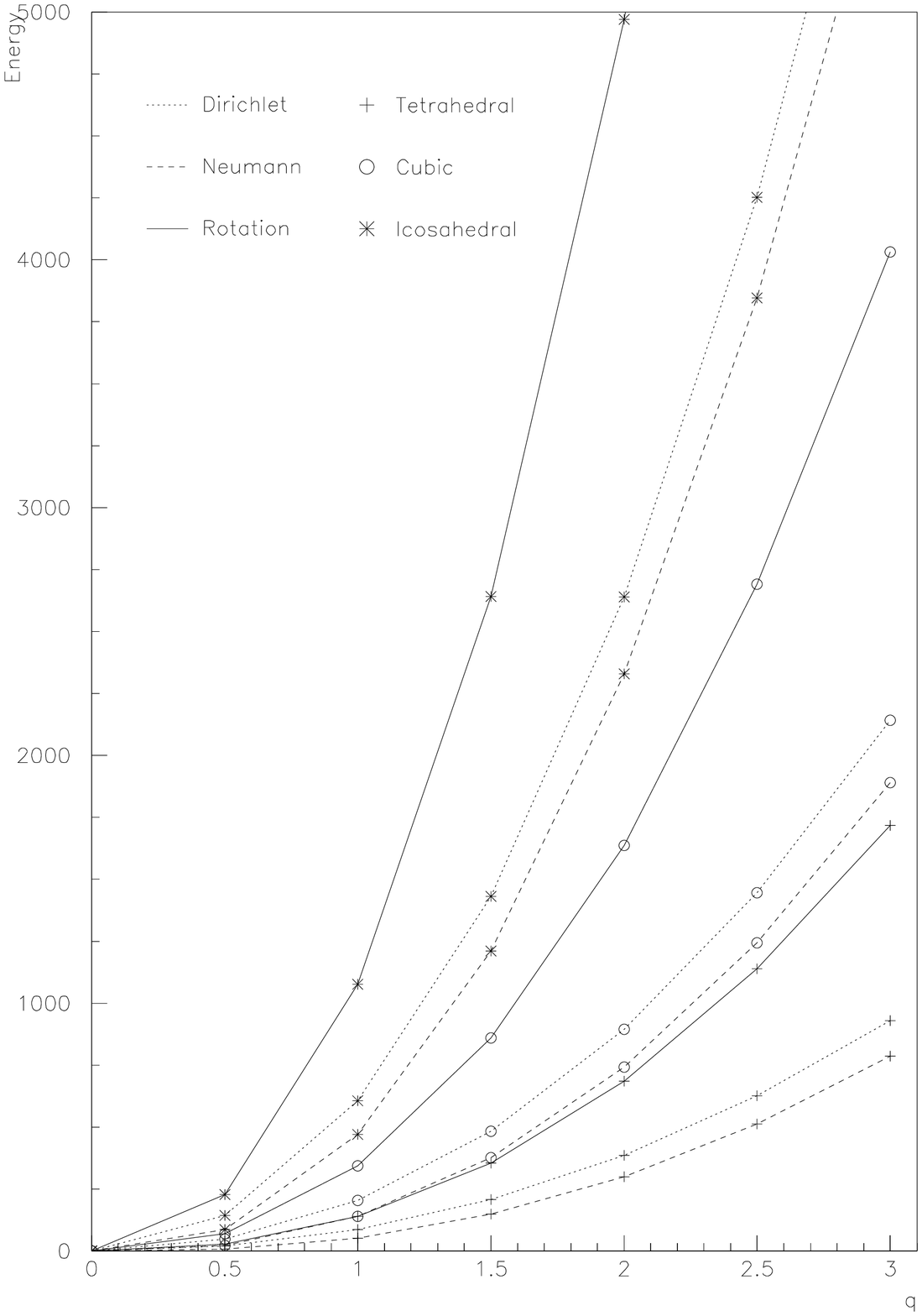}

Fig.4  Vacuum energies for groups $T$, $O$ and $Y$ as functions of
$\overline q'$.
\vskip20truept
The graphs with $\overline q=q=0$ are exactly the same as those
calculated in
[\pref{Dow1}]. To plot the graphs in this case we have had
to remove the divergence in the first Barnes zeta function derivative in
(\peq{mon:zDp0}). This is due to a zero mode appearing in the spectrum of the
operator $H_{S^2}$. The divergence is logarithmic and can be removed by
defining the so called $\Gamma$-modular form $\rho_2$ defined
by [\pref{Barnesa,Dow1}],
$$
\lim_{a\to 0}\zeta'_2(0,a|d_1,d_2)=-\ln a-\ln\rho_2.
$$
The leading term $-\ln a$ on the right hand side is the divergent zero mode
contribution which must be removed. The form $\rho_2$ has been calculated in
terms of the multiple gamma function by Barnes [\pref{Barnesa,Barnesb}]. The
$\Gamma$-modular form can be more easily calculated by using the
simply proved relation
$$
\zeta_2(s,a|d_1,d_2)=a^{-s}+d_1^{-s}\zeta_H(s,1+\frac{a}{d_1})+
d_2^{-s}\zeta_H(s,1+\frac{a}{d_2})+\zeta_2(s,a+d_1+d_2|d_1,d_2).
$$
Differentiating this with respect to $s$ at $s=0$, and taking the limit $a\to
0$ leaves the expression
$$
-\ln\rho_2=\half\ln(d_1 d_2)-\ln(2\pi)+\zeta'_2(0,d_0+1|d_1,d_2).
$$

For $q>0$, the zero mode that we have just removed for $q=0$ is still almost
zero. This is the reason why the vacuum energy for the Dirichlet zeta
function increases more rapidly with $\overline q'$ than in the Neumann case.

\section{\bf 11. Theory for a spherical slice}

By a spherical slice of width $\beta$ we mean the space
$S_\beta=\{(\theta,\phi)|\theta\in[0,\pi]\;,\ \phi\in[0,\beta]\}$ with the
points $(\theta,0)$ and $(\theta,\beta)$ identified (here $(\theta,\phi)$ are
the spherical polar coordinates on the sphere which shall be used throughout
this section). One might also term this space a `periodic lune'.

The starting point for the monopole theory on $S_\beta$ is the
solution to the explicit differential equation (\peq{mon:diff}) with
arbitrary
$q$. As mentioned in the discussion surrounding the differential equation the
solutions are characterised by integers $u,v$, and are given by equations
(\peq{mon:harm1}), (\peq{mon:harm2}). We shall now show that it is possible
 to
define consistent monopole theories on $S_\beta$ corresponding to the
rotational, Dirichlet and Neumann cases already given. The rotational case is
considered first.

Define $k$ by $k=2\pi/\beta$ and let $\widehat\gamma$ denote a rotation by
angle $\beta$ about the $z$ axis. Due to the identification of points in
$S_\beta$ at
$\phi=0,\beta$, we will proceed conventionally and take the wave
function to be single valued \ie to have period $\beta$. Also, for
rapidity, the monopole charge quantisation will be obtained by
requiring the connecting function $\exp(i2q\phi)$ to have period
$\be$. This yields

\begin{ignore}
if $\varphi(\theta,\phi)$ is a monopole wave function then
$\varphi$ must satisfy the equation
$$
e^{-iq\Lambda_{\hat\gamma}(\theta,\beta)}\varphi(\theta,0)=
a(\widehat\gamma)\varphi(\theta,\beta).
$$
This is equation (\peq{period}) specialised to the boundary of
$S_\beta$. (For the particular $\widehat\ga$,
$\Lambda_{\hat\gamma}(\theta,\beta)$ is independent of $\th$.)

If we insist that $\varphi$ does not vanish at the fixed points
$z=\pm 1$ of $\hat\gamma$, then we end up with the consistency equations
corresponding to (\peq{fp}),
$$
e^{-iq\beta}\varphi(0,\phi)=a(\widehat\gamma)\varphi(0,\phi)\;,\quad
e^{+iq\beta}\varphi(\pi,\phi)=a(\widehat\gamma)\varphi(\pi,\phi)
\eql{mon:moncon2}$$
Matching the two values of $a(\widehat\gamma)$ gives the values of the
 monopole charge, $q$, as
\end{ignore}

$$
q=\frac{k}{2}n\;,\quad n\in\oZ\,.
\eql{mon:quant5}$$
Corresponding to these values of $q$ we find $a(\widehat\gamma)=(-1)^n$.
If $k$ is
an integer then $S_\beta$ is a fundamental domain for the group $C_k$
(generated by $\widehat\gamma$), and the relation $|C_k|=k$ gives an
equivalence
between the quantisation conditions (\peq{quant3}) and (\peq{mon:quant5}).
Thus we see that the quantisation condition for arbitrary $k$ is a
generalisation of the previous theory for $C_k$ (with integer $k$).
From single valuedness,
$
\varphi(\theta,\beta)=\varphi(\theta,0),
$
for the monopole harmonics (\peq{mon:harm2}), we see from the explicit
$\phi$
dependence that,
$$
m=jk-q\;,\quad j\in\oZ\,.
\eql{mon:m}$$
In this more general setting it would appear that we can always choose $n$ to
be odd, whereas before this was only possible for $k$ odd also. From
(\peq{mon:harm2})) we find the possible values of $l$,
$$
l=u+\frac{k}{2}(|j|+|n-j|)\;,\quad u\in\oZ^+\;,\ j\in\oZ\,.
$$
The eigenvalues $\lambda$ appearing in the mode equation (\peq{mon:diff}) are
still given by $l(l+1)-q^2$. As in section 7 we shall calculate the
linear heat kernel for the eigenvalues $(l+1/2)$.

Using the values of $q$, $l$, and $m$ above gives the sum form for the linear
heat kernel analogous to (\peq{linker2}), with (\peq{mon:harm2}),
$$
\widetilde{K}^q_k(\tau)=e^{-\tau/2}\sum_{u=0}^\infty\sum_{j=-\infty}^\infty
e^{-\tau(u+k(|j|+|j+n|)/2}\,.
\eql{mon:cycsum5}$$
Evaluating this sum gives exactly the result found before, equation
(\peq{lingen}), except that now $k$ is not an integer, and $q$ need not
be an integer multiple of $k$. A corollary of this result is that
(\peq{lingen}) is also valid for the case when $\Gamma=C_k$ with $k$ and
$2q$ odd integers. The zeta function defined by equation (\peq{genzeta}) is
also correct for $d_1=1$, $d_2=k$ and hence we conclude that the vacuum
energies calculated in the previous section  are valid for {\it arbitrary}
 $k$.

For the Dirichlet and Neumann theories defined in section 8 we
consider the slice $S_{\beta/2}$ and two reflection planes $P_0$ and $P_1$
which leave $\phi=0$ and $\phi=\beta/2$ invariant respectively. Following the
same arguments as in section 8 gives the values of the monopole
charge
$$
q=k n'\,,
\eql{mon:quant6}$$
which is equivalent to equation (\peq{quant2}). Taking $\gamma'$ in
(\peq{reflmods}) to be a reflection in $P_0$ and using (\peq{reflact}) gives the
modes $W^{(l)}_{qm}$ as the combinations,
$$
W^{(l)}_{qm}(\theta,\phi)=Y^{(l)}_{qm}(\theta,\phi)
+a(\si)Y^{(l)}_{q\,-m}(\theta,\phi)
$$
Here $Y^{(l)}_{qm}$ are those defined in (\peq{mon:harm1}) with the string
along the
negative $z$ axis, and $a(\si)=\pm 1$ as before (the reflection in $P_0$ is
expressed simply as $\phi\to -\phi$).

The $\phi$ dependent part of the modes $W^{(l)}_{qm}$ is given by
$$
W^{(l)}_{qm}\sim e^{iq\phi}\times\left\{\cos(m\phi)\;,\quad a(\si)=+1 \atop
\sin(m\phi)\;,\quad a(\si)=-1 \right. \,.
$$
Reflection in $P_0$ is equivalent to $\phi\to -\phi$ and we see explicitly
that this transforms $W^{(l)}_{qm}(\theta,0)$ into
$a(\si)W^{(l)}_{-qm}(\theta,0)$ as required. Reflection in $P_1$ is
equivalent to $\phi\to\beta-\phi$ and results in the condition on $m$,
$$
m=k j\;,\quad j\in\oZ^+\,.
\eql{mon:m2}$$
For the Dirichlet case $a(\si)=-1$ the $m=0$ mode is in fact zero and has to
be
removed. Comparing (\peq{mon:m2}) with (\peq{mon:m}) (with $n=2n'$ even) we
see
that the only difference between the rotation and reflection cases is that in
the reflection case the values of $m$ are restricted to positive integers.

The linear heat kernel for the Neumann case $a(\si)=+1$ is given by
$$
\widetilde{K}^q_N(\tau)=e^{-\tau/2}\sum_{u=0}^\infty\sum_{j=0}^\infty
e^{-\tau(u+k(|j+n'|+|j-n'|)/2}\,.
\eql{mon:cycsum6}$$
Explicit evaluation of the sum gives exactly the heat kernel as before,
equation (\peq{lincycl}). The Dirichlet case involves subtracting the $j=0$
term from (\peq{cyclrefld}) and yields the previous expression
(\peq{dihedrefld}). Thus the zeta functions are given exactly as before and
we
conclude that the vacuum energy calculated for $C_k$ is in fact valid for all
$k$.

To extend the results of $D_k$ to arbitrary $k$ we consider the slice of the
upper hemisphere $S'_\beta=\{(\theta,\phi)|\theta\in[0,\pi/2]\;,\
\phi\in[0,\beta]\}$ with again $k=2\pi/\beta$. This is a fundamental domain
for
$D_k$ when $k$ is an integer.  The theory then follows as for the $C_k$
extension above, but we must include in this case a rotation about the $x$
axis
by angle $\pi$. This rotation can be thought of as a reflection in the plane
$P_0$ followed by a reflection in the $x-y$ plane, which we call $P_2$. The
theory for $S_\beta$ above is adapted to the reflection $P_0$ and thus we see
that $P_2$ is the essential extra detail here.

The reflection $P_2$ is equivalent to the transformation
$\theta\to\pi-\theta$
which does not affect the $\phi$ dependence of the modes
$Y^{(l)}_{qm}(\theta,\phi)$. Since the extension to arbitrary $k$ is entirely linked
with the $\phi$ dependence, we conclude that all the heat kernels, zeta
functions and vacuum energies for $D_k$ can be extended to arbitrary $k$. The
values of the monopole charge in the reflection case are calculated using the
theory of section 8 as
$$
q=2k n'\;,\quad n'\in\oZ,
$$
which is the generalisation of (\peq{quant2}) with $|\Gamma'|=4k$.

\section{\bf 12. Summary and discussion}

We have thoroughly adapted Dirac's monopole theory to the
orbifold, $\rS^2/\Ga$, for the cases that $\Ga$ contains only
rotations and when $\Ga$ is generated by reflections. In the
former case we imposed rotational (periodic) boundary conditions on the
monopole solutions. In the latter we had a choice of boundary
conditions defined so as to reproduce Dirichlet and Neumann
conditions for no monopole charge, $q=0$. We found that it was the
monopole charge $\overline q=q/|\Ga|$ through $\rS^2/\Ga$ that was
Dirac quantised with $2\overline q \in\oZ$.

After all the formalities of the theory had been tidied we
explicitly calculated the vacuum energies on the orbifolds
$\rS^2/\Ga$ and $\oR\times \rS^2/\Ga$. Formal expressions are
given for the generalisation to the spaces
$\oR^{2n}\times\rS^2/\Ga$ and $\oR\times\oR^{2n}\times\rS^2/\Ga$.
Finally we provided an extension of the monopole theory to
arbitrary slices of the sphere and hemisphere. In this case the
flux through the spherical region is still quantised, although now
the overall monopole charge $q$ is not, in general, an integer or
half odd-integer.

We feel that the scalar theory has been developed essentially to
its analytical limit on the factored sphere. The next step would
be the extension to $\oR^3/\Ga$ for $\Ga$ a reflection or rotation
group. This requires modes of the full Hamiltonian which are given
by $$ Y^{(l)}_{qm}(\th,\phi)J_\nu(kr)\sqrt{k/r},\quad
\nu=\sqrt{(l+1/2)^2-q^2} $$ with eigenvalues $k^2$. Since the
radial dependence does not involve $m$, the underlying facts of
the theory (modes on factored space, charge quantisation \etc) are
the same as in the spherical case. However the heat-kernel
calculation, and hence the \zf, is completely different. Due to
the complicated index, $\sqrt{(l+1/2)^2-q^2}$, closed forms do not
seem possible and asymptotic methods are needed. One could always
arbitrarily add a term $q^2/r$ to the (total) Hamiltonian and then
a closed form would exist. This fact suggests that there is some
significance to this modification.

The spinor theory on the factored sphere $\rS^2/\Ga$ has been
considered by Chang [\pref{Chang}]. He found a consistent theory
only for $\Ga=C_k$ with $k$ odd. For $q\ne0$ we claim that the
same restriction still holds. This follows from the lack of half
odd-integral solutions to the scalar monopole problem for $\Ga\ne
C_k$.

A possible extension of the scalar calculation would be to
consider if (high temperature) Bose-Einstein condensation occurs.
The general theory has been laid down by Toms [\pref{Toms}] See also
Kirsten and Toms [\pref{KandT}].
Basically, all that is required is to ensure that the \zf\ for the
theory, and its derivative, are finite at zero as the chemical
potential approaches a critical value. On the two-sphere we can
use the calculations of $\ze(-n)$ and $\ze'(-n$ given in section 9
to study the theory on $\oR^{2n}\times\rS^2$. In fact we could also
discuss $\oR^{2n}\times\rS^d$.

\section{\bf Acknowledgments}
I would like to thank Chris Isham for useful discussions and the
Theory Group at Imperial College for hospitality while this
work was carried out under the EPSRC grant No. GR/M08714.

\section{\bf References}
\vskip 5truept
\begin{putreferences}
\ref{APS}{Atiyah,M.F., V.K.Patodi and I.M.Singer: Spectral asymmetry and
Riemannian geometry \mpcps{77}{75}{43}.}
\ref{AandT}{Awada,M.A. and D.J.Toms: Induced gravitational and gauge-field
actions from quantised matter fields in non-abelian Kaluza-Klein thory
\np{245}{84}{161}.}
\ref{BandI}{Baacke,J. and Y.Igarishi: Casimir energy of confined massive
quarks \prD{27}{83}{460}.}
\ref{Barnesa}{Barnes,E.W.
{\it Trans. Camb. Phil. Soc.} {\bf 19} (1903) 374.}
\ref{Barnesb}{Barnes,E.W. {\it Trans. Camb. Phil. Soc.} {\bf 19} (1903) 426.}
\ref{Barv}{Barvinsky,A.O. Yu.A.Kamenshchik and I.P.Karmazin: One-loop
quantum cosmology \aop {219}{92}{201}.}
\ref{BandM}{Beers,B.L. and Millman, R.S. :The spectra of the
Laplace-Beltrami
operator on compact, semisimple Lie groups. \ajm{99}{1975}{801-807}.}
\ref{BandH}{Bender,C.M. and P.Hays: Zero point energy of fields in a
confined volume \prD{14}{76}{2622}.}
\ref{BandL}{Biedenharn,L.C. and Louck,J.D. {\it The Racah-Wigner
Algebra in Quantum Theory}, (Addison-Wesley,Reading,Mass,1981).}
\ref{BBG}{Bla\v zi\' c,N., Bokan,N. and Gilkey,P.B.: Spectral geometry
of the
form valued Laplacian for manifolds with boundary \ijpam{23}{92}{103-120}}
\ref{BEK}{Bordag,M., E.Elizalde and K.Kirsten: { Heat kernel
coefficients of the Laplace operator on the D-dimensional ball},
\jmp{37}{96}{895}.}
\ref{BGKE}{Bordag,M., B.Geyer, K.Kirsten and E.Elizalde,: { Zeta function
determinant of the Laplace operator on the D-dimensional ball},
\cmp{179}{96}{215}.}
\ref{BKD}{Bordag,M., K.Kirsten,K. and Dowker,J.S.: Heat kernels and
functional determinants on the generalized cone \cmp{182}{96}{371}.}
\ref{Branson}{Branson,T.P.: Conformally covariant equations on differential
forms \cpde{7}{82}{393-431}.}
\ref{BandG2}{Branson,T.P. and Gilkey,P.B. {\it Comm. Partial Diff. Eqns.}
{\bf 15} (1990) 245.}
\ref{BGV}{Branson,T.P., P.B.Gilkey and D.V.Vassilevich {\it The Asymptotics
of the Laplacian on a manifold with boundary} II, hep-th/9504029.}
\ref{BandS}{Brink,D.M. and Satchler,G.R. {\it Angular Momentum}
2nd ed. Clarendon Press, Oxford, 1968.}
\ref{BCZ1}{Bytsenko,A.A, Cognola,G. and Zerbini, S. : Quantum fields in
hyperbolic space-times with finite spatial volume, hep-th/9605209.}
\ref{BCZ2}{Bytsenko,A.A, Cognola,G. and Zerbini, S. : Determinant of
Laplacian on a non-compact 3-dimensional hyperbolic manifold with finite
volume, hep-th /9608089.}
\ref{CandH2}{Camporesi,R. and Higuchi, A.: Plancherel measure for $p$-forms
in real hyperbolic space, \jgp{15}{94}{57-94}.}
\ref{CandH}{Camporesi,R. and A.Higuchi {\it On the eigenfunctions of the
Dirac operator on spheres and real hyperbolic spaces}, gr-qc/9505009.}
\ref{CandW}{Candelas,P. and Weinberg,S. \np{257}{84}{397}}
\ref{CEJM}{Chamblin,A.,Emparan,R.,Johnson,C.V. and Myers,R.C.
 {\it Large N Phases. Gravitational Instantons and the Nuts and
Bolts of AdS Holography} \break hep-th/9808177.}
\ref{ChandD}{Chang, Peter and Dowker,J.S. \np{395}{93}{407}.}
\ref{ChandD2}{Chang, Peter and Dowker,J.S. \prD{}{}{}.}
\ref{Chang}{Chang,Peter, PhD thesis, University of Manchester,
1993.}
\ref{cheeg1}{Cheeger, J.: Spectral Geometry of Singular Riemannian Spaces.
\jdg {18}{83}{575}.}
\ref{cheeg2}{Cheeger,J.: Hodge theory of complex cones {\it Ast\'erisque}
{\bf 101-102}(1983) 118-134}
\ref{Chou}{Chou,A.W.: The Dirac operator on spaces with conical
singularities and positive scalar curvature, \tams{289}{85}{1-40}.}
\ref{CandT}{Copeland,E. and Toms,D.J.: Quantized antisymmetric
tensor fields and self-consistent dimensional reduction
in higher-dimensional spacetimes, \break \np{255}{85}{201}}
\ref{CS}{ {\it The Formation and Evolution of Cosmic strings}
edited by G.W.Gibbons, S.W.Hawking and T.Vachaspati (Cambridge
University Press, Cambridge, 1990.}
\ref{DandH}{D'Eath,P.D. and J.J.Halliwell: Fermions in quantum cosmology
\prD{35}{87}{1100}.}
\ref{cheeg3}{Cheeger,J.:Analytic torsion and the heat equation. \aom{109}
{79}{259-322}.}
\ref{Cook}{Cook, A.W. PhD Thesis, University of Manchester, 1996.}
\ref{DandE}{D'Eath,P.D. and G.V.M.Esposito: Local boundary conditions for
Dirac operator and one-loop quantum cosmology \prD{43}{91}{3234}.}
\ref{DandE2}{D'Eath,P.D. and G.V.M.Esposito: Spectral boundary conditions
in one-loop quantum cosmology \prD{44}{91}{1713}.}
\ref{CrandD}{Critchley,R. and Dowker,J.S. \jpa{14}{81}{1943};
{\it ibid} {\bf 15} (1982) 157.}
\ref{Dow1}{Dowker,J.S. \cmp{162}{94}{633}.}
\ref{Dow8}{Dowker,J.S. {\it Robin conditions on the Euclidean ball}
MUTP/95/7; hep-th\break/9506042. {\it Class. Quant.Grav.} to be published.}
\ref{Dow9}{Dowker,J.S. {\it Oddball determinants} MUTP/95/12;
hep-th/9507096.}
\ref{Dow10}{Dowker,J.S. {\it Spin on the 4-ball},
hep-th/9508082, {\it Phys. Lett. B}, to be published.}
\ref{dow11}{Dowker,J.S. {\it Vacuum energy on the squashed 3-sphere}
in {\it Quantum Gravity} ed by S.C.Christensen 1984 IOP, Bristol.}
\ref{DandA2}{Dowker,J.S. and J.S.Apps, {\it Functional determinants on
certain domains}. To appear in the Proceedings of the 6th Moscow Quantum
Gravity Seminar held in Moscow, June 1995; hep-th/9506204.}
\ref{DandK}{Dowker,J.S. and Kennedy,G. \jpa {11}{78}{895}.}
\ref{DABK}{Dowker,J.S., Apps,J.S., Bordag,M. and Kirsten,K.: Spectral
invariants for the Dirac equation with various boundary conditions
{\it Class. Quant.Grav.} to be published, hep-th/9511060.}
\ref{EandR}{E.Elizalde and A.Romeo : An integral involving the
generalized zeta function, {\it International J. of Math. and
Phys.} {\bf13} (1994) 453.}
\ref{ELV2}{Elizalde, E., Lygren, M. and Vassilevich, D.V. : Zeta function
for the laplace operator acting on forms in a ball with gauge boundary
conditions. hep-th/9605026}
\ref{ELV1}{Elizalde, E., Lygren, M. and Vassilevich, D.V. : Antisymmetric
tensor fields on spheres: functional determinants and non-local
counterterms, \jmp{}{96}{} to appear. hep-th/ 9602113}
\ref{Kam2}{Esposito,G., A.Y.Kamenshchik, I.V.Mishakov and G.Pollifrone:
Gravitons in one-loop quantum cosmology \prD{50}{94}{6329};
\prD{52}{95}{3457}.}
\ref{Erdelyi}{A.Erdelyi,W.Magnus,F.Oberhettinger and F.G.Tricomi {\it
Higher Transcendental Functions} Vol.I McGraw-Hill, New York, 1953.}
\ref{Esposito}{Esposito,G.: { Quantum Gravity, Quantum Cosmology and
Lorentzian Geometries}, Lecture Notes in Physics, Monographs, Vol. m12,
Springer-Verlag, Berlin 1994.}
\ref{Esposito2}{Esposito,G. {\it Nonlocal properties in Euclidean Quantum
Gravity}. To appear in Proceedings of 3rd. Workshop on Quantum Field Theory
under the Influence of External Conditions, Leipzig, September 1995;
gr-qc/9508056.}
\ref{EKMP}{Esposito G, Kamenshchik Yu A, Mishakov I V and Pollifrone G.:
One-loop Amplitudes in Euclidean quantum gravity.
\prD{52}{96}{3457}.}
\ref{ETP}{Esposito,G., H.A.Morales-T\'ecotl and L.O.Pimentel {\it Essential
self-adjointness in one-loop quantum cosmology}, gr-qc/9510020.}
\ref{Fierz}{Fierz,M. \hpa{17}{44}{27}.}
\ref{FORW}{Forgacs,P., L.O'Raifeartaigh and A.Wipf: Scattering theory,
U(1) anomaly and index theorems for compact and non-compact manifolds
\np{293}{87}{559}.}
\ref{FandH}{Frenkel,A. and Hrask\'o,P. \aop{105}{77}{288}.}
\ref{GandM}{Gallot S. and Meyer,D. : Op\'erateur de coubure et Laplacian
des formes diff\'eren-\break tielles d'une vari\'et\'e riemannienne
\jmpa{54}{1975}{289}.}
\ref{Gibb}{Gibbons, G.W. \aop {125}{80}{98}.}
\ref{Gilkey1}{Gilkey, P.B, Invariance theory, the heat equation and the
Atiyah-Singer index theorem, 2nd. Edn., CRC Press, Boca Raton 1995.}
\ref{Gilkey2}{Gilkey,P.B.:On the index of geometric operators for
Riemannian manifolds with boundary \aim{102}{93}{129}.}
\ref{Gilkey3}{Gilkey,P.B.: The boundary integrand in the formula for the
signature and Euler characteristic of a manifold with boundary
\aim{15}{75}{334}.}
\ref{GandP}{Greub,W. and Petry,H,-R. \jmp{16}{75}{1347}.}
\ref{Grubb}{Grubb,G. {\it Comm. Partial Diff. Eqns.} {\bf 17} (1992)
2031.}
\ref{GandS1}{Grubb,G. and R.T.Seeley \cras{317}{1993}{1124}; \invm{121}{95}
{481}.}
\ref{GandS}{G\"unther,P. and Schimming,R.:Curvature and spectrum of compact
Riemannian manifolds, \jdg{12}{77}{599-618}.}
\ref{Hitch}{Hitchin,N. \aim {14}{74}{1}.}
\ref{HFP}{Hu,B-.L, Fulling,S.A. and Parker, L. \prD{8}{73}{2377}.}
\ref{Hu}{Hu,B-.L. \prD{8}{73}{1048}.}
\ref{Horvathy}{Horvathy,P.\ijtp {20}{81}{697}.}
\ref{HHP}{Hawking,S.W, Hunter,C.J. and Page, D.N {\it Nut Charge,
Anti-de Sitter Space and Entropy} hep-th/9809035.}
\ref{IandT}{Ikeda,A. and Taniguchi,Y.:Spectra and eigenforms of the
Laplacian
on $S^n$ and $P^n(C)$. \ojm{15}{1978}{515-546}.}
\ref{IandK}{Iwasaki,I. and Katase,K. :On the spectra of Laplace operator
on $\La^*(S^n)$ \pja{55}{79}{141}.}
\ref{JandJ}{Janner,A.and Janssen,T. \physica {53}{71}{1}.}
\ref{JandK}{Jaroszewicz,T. and P.S.Kurzepa: Polyakov spin factors and
Laplacians on homogeneous spaces \aop{213}{92}{135}.}
\ref{Kam}{Kamenshchik,Yu.A. and I.V.Mishakov: Fermions in one-loop quantum
cosmology {47}{93}{1380}.}
\ref{KandM}{Kamenshchik,Yu.A. and I.V.Mishakov: Zeta function technique for
quantum cosmology {\it Int. J. Mod. Phys.} {\bf A7} (1992) 3265.}
\ref{KandT}{Kirsten,K. and Toms,D.J. {\it Bose-Einstein
condensation in arbitrarily shaped cavities}, cond-mat/9810098.}
\ref{KandC}{Kirsten,K. and Cognola.G,: { Heat-kernel coefficients and
functional determinants for higher spin fields on the ball} \cqg{13}{96}
{633-644}.}
\ref{Levitin}{Levitin,M.: { Dirichlet and Neumann invariants for Euclidean
balls}, {\it Diff. Geom. and its Appl.}, to be published.}
\ref{Lindel}{Lindel\"of.E. {\it Le Calcul des R\'esidus},
Gauthier-Villars, Paris, 1905.}
\ref{Luck}{Luckock,H.C.: Mixed boundary conditions in quantum field theory
\jmp{32}{91}{1755}.}
\ref{MandL}{Luckock,H.C. and Moss,I.G,: The quantum geometry of random
surfaces and spinning strings \cqg{6}{89}{1993}.}
\ref{Ma}{Ma,Z.Q.: Axial anomaly and index theorem for a two-dimensional
disc
with boundary \jpa{19}{86}{L317}.}
\ref{Mcav}{McAvity,D.M.: Heat-kernel asymptotics for mixed boundary
conditions \cqg{9}{92}{1983}.}
\ref{MOS1966}{Magnus,W., Oberhettinger,F. and Soni, R.P., {\it
Formulas and theorems for the special functions of mathematical
physics}, 3rd. ed. (Springer-Verlag, Berlin, 1966.}
\ref{MandV}{Marachevsky,V.N. and D.V.Vassilevich {\it Diffeomorphism
invariant eigenvalue \break problem for metric perturbations in a bounded
region}, SPbU-IP-95, \break gr-qc/9509051.}
\ref{Milton}{Milton,K.A.: Zero point energy of confined fermions
\prD{22}{80}{1444}.}
\ref{Meyer}{Meyer,B.\cjm{6}{54}{135}.}
\ref{MandS}{Mishchenko,A.V. and Yu.A.Sitenko: Spectral boundary conditions
and index theorem for two-dimensional manifolds with boundary
\aop{218}{92}{199}.}
\ref{Moss}{Moss,I.G.: Boundary terms in the heat-kernel expansion
\cqg{6}{89}{759}.}
\ref{MandP}{Moss,I.G. and S.J.Poletti: Conformal anomaly on an Einstein
space
with boundary \pl{B333}{94}{326}.}
\ref{MandP2}{Moss,I.G. and S.J.Poletti \np{341}{90}{155}.}
\ref{NandOC}{Nash, C. and O'Connor,D.J.: Determinants of Laplacians, the
Ray-Singer torsion on lens spaces and the Riemann zeta function
\jmp{36}{95}{1462}.}
\ref{NandS}{Niemi,A.J. and G.W.Semenoff: Index theorem on open infinite
manifolds \np {269}{86}{131}.}
\ref{NandT}{Ninomiya,M. and C.I.Tan: Axial anomaly and index thorem for
manifolds with boundary \np{245}{85}{199}.}
\ref{norlund2}{N\"orlund~N. E.:M\'emoire sur les polynomes de Bernoulli.
\am {4}{21} {1922}.}
\ref{Okada}{Okada,Y. \cqg{}{86}{221}.}
\ref{OandT}{Opechowski,W. and Tam, W.G. \physica {42}{69}{529}.}
\ref{Peierls}{Peierls,R. \zfp {80}{33}{763}.}
\ref{Poletti}{Poletti,S.J. \pl{B249}{90}{355}.}
\ref{RandT}{Russell,I.H. and Toms D.J.: Vacuum energy for massive forms
in $R^m\times S^N$, \cqg{4}{86}{1357}.}
\ref{RandS}{R\"omer,H. and P.B.Schroer \pl{21}{77}{182}.}
\ref{SandS}{Shen, T.C. and Sobczyk,J. \prD{36}{87}{397}.}
\ref{SHC}{Shen,T.C., Hu,B.L. and O'Connor,D.J. \prD{31}{85}{2401}.}
\ref{SandV}{Shtykov,N. and Vassilevich, D.V. \jpa {28}{95}{L37}.}
\ref{Snia}{Sniatycki,J.\jmp {15}{74}{619}.}
\ref{Tamm}{Tamm,I. \zfp{71}{31}{141}.}
\ref{Tam}{Tam,W.G. \physica {}{}{}.}
\ref{tetra}{Dowker,J.S. \jmp {28}{87}{33}.}
\ref{Toms}{Toms,D.J. \prD {50}{94}{6457}.}
\ref{Trautman}{Trautman,A.: Spinors and Dirac operators on hypersurfaces
\jmp{33}{92}{4011}.}
\ref{Vass}{Vassilevich,D.V.{Vector fields on a disk with mixed
boundary conditions} gr-qc /9404052.}
\ref{Voros}{Voros,A.:
Spectral functions, special functions and the Selberg zeta function.
\cmp{110}{87}439.}
\ref{WandY}{Wu,T.T. and Yang,C.N. \np{107}{76}{365}; \prD{16}{77}{1018}.}
\ref{Ray}{Ray,D.B.: Reidemeister torsion and the Laplacian on lens
spaces \aim{4}{70}{109}.}
\ref{McandO}{McAvity,D.M. and Osborn,H. Asymptotic expansion of the heat kernel
for generalised boundary conditions \cqg{8}{91}{1445}.}
\ref{AandE}{Avramidi,I. and Esposito,G. Heat kernel asymptotics with
generalised boundary conditions, hep-th/9701018.}
\ref{MandS}{Moss,I.G. and Silva P.J., Invariant boundary conditions for
gauge theories gr-qc/9610023.}
\ref{barv}{Barvinsky,A.O.\pl{195B}{87}{344}.}
\ref{krantz}{Krantz,S.G. Partial Differential Equations and Complex
Analysis (CRC Press, Boca Raton, 1992).}
\ref{treves}{Treves,F. Introduction to Pseudodifferential and Fourier
Integral Operators,\break Vol.1, (Plenum Press,New York,1980).}
\ref{EandS}{Egorov,Yu.V. and Shubin,M.A. Partial Differential Equations
(Springer-Verlag, Berlin,1991).}
\ref{AandS}{Abramowitz,M. and Stegun,I.A. Handbook of Mathematical
Functions (Dover, New York, 1972).}
\ref{ACNY}{Abouelsaood,A., Callan,C.G., Nappi,C.R. and Yost,S.A.\np{280}
{87}{599}.}
\ref{BGKE}{Bordag,M., B.Geyer, K.Kirsten and E.Elizalde, { Zeta function
determinant of the Laplace operator on the D-dimensional ball},
\cmp{179}{96}{215}.}

\end{putreferences}
 \bye